\documentclass[conference,pageno,bookmarks=false,breaklinks=true,colorlinks,linkcolor=black,citecolor=blue,urlcolor=black]{IEEETran}

\usepackage[normalem]{ulem}
\usepackage{makecell}
\usepackage{cite}
\usepackage{amsmath,amssymb,amsfonts}
\usepackage{mathptmx} % This is Times font
\usepackage{comment}
\usepackage{bm}
\usepackage{algpseudocode}
\usepackage{algorithm, algorithmicx}%
\usepackage{graphicx}
\usepackage{textcomp}
\usepackage{multirow}
\usepackage{fancyhdr}
\usepackage{MnSymbol}

\title{\name: An Optimization Framework for Machine-Discovered Network Topologies}
\author{Conor Green and Mithuna Thottethodi\\
Elmore Family School of Electrical and Computer Engineering, Purdue University\\
\{green456,mithuna\}@purdue.edu
}

\date{}

\usepackage{ifthen}
\usepackage{calc}
\usepackage{pifont}
\usepackage{color}
\usepackage{fancyhdr}
\usepackage{xspace}
\usepackage{url}
\usepackage{xkeyval}
\usepackage{multirow}
\usepackage[table]{xcolor} % Needed for stripetabular environment
\definecolor{lightgray}{gray}{0.9}

\newcounter{hours}
\newcounter{minutes}

\newcommand{\etal}{{\it et al.}\xspace}

\newcommand{\name}{{NetSmith}\xspace}

\newboolean{timeofmake}
\setboolean{timeofmake}{false}

\newcommand{\dontinclude}[1]{ }

\newcommand{\putsec}[2]{\vspace{-0.1in}\section{#2}\label{sec:#1}\vspace{0.0in}}
\newcommand{\putsubsec}[2]{\vspace{-0.05in}\subsection{#2}\label{sec:#1}\vspace{0.0in}}

\newcommand{\tabput}[3]{
\begin{table}[t]
\caption{#3 \label{tab:#1}}
\vspace{-0.1in}
\begin{center}
{
#2
}
\end{center}
\vspace{-0.15in}
\end{table}
}

\newcommand{\figput}[4][1.0\linewidth]{
\begin{figure}[t]
\begin{minipage}{\linewidth}
\footnotesize 
\begin{center}
\includegraphics[width=#1]{figures/#2}
\end{center}
\vspace{-0.05in}

\caption{#4 \label{fig:#2}}

\vspace{-0.05in}

\end{minipage}
\end{figure}
}

\newcommand{\figputV}[6][1.0\linewidth]{
\begin{figure}[t]
\begin{minipage}{\linewidth}
\footnotesize
\begin{tabular}{c}
\includegraphics[width=#1]{figures/#2}\\
(a) {#3}\\
\includegraphics[width=#1]{figures/#4}\\
(b) {#5}\\
\end{tabular}
\caption{#6 \label{fig:#2}}
\end{minipage}
\end{figure}
}

\newcommand{\figputW}[4][\linewidth]{
\begin{figure*}
\begin{minipage}{\linewidth}
\footnotesize 
\begin{center}
\includegraphics[width=#1]{figures/#2}
\end{center}
\vspace{-0.2in}
\caption{#4 \label{fig:#2}}
\end{minipage}

\end{figure*}
}

\newcommand{\figref}[1]{Figure~\ref{fig:#1}}
\newcommand{\tabref}[1]{Table~\ref{tab:#1}}
\newcommand{\secref}[1]{Section~\ref{sec:#1}}

\begin{document}

\maketitle
\thispagestyle{empty}

\pagestyle{plain}

\begin{abstract} 

Over the past few decades, network topology design for general purpose, shared memory multicores has been primarily driven by human experts who use their insights to arrive at network designs that balance the competing goals of performance requirements (e.g., latency, bandwidth) and cost constraints (e.g., router radix, router counts). On the other hand, there have been automatic NoC synthesis methods for SoCs to optimize for application-specific communication and objectives such as resource usage or power. 
Unfortunately, these techniques do not lend themselves to the general-purpose context, where directly applying these previous NoC synthesis techniques in the general-purpose context yields poor results, even worse than expert-designed networks. 
We design and develop an automatic network design methodology -- \name -- to design networks for general-purpose, shared memory multicores that comprehensively outperform expert-designed networks.

We employ \name in the context of interposer networks for chiplet-based systems where there has been significant recent work on network topology design (e.g., Kite, Butter Donut, Double Butterfly). \name generated topologies are capable of achieving significantly higher throughput (50\% to 75\% higher) while also reducing average hop count by 8\% to 13.5\%) than previous expert-designed and synthesized networks.
Full system simulations using PARSEC benchmarks demonstrate that the improved network performance translates to improved application performance with up to 11\% mean speedup over previous NoI topologies.
\end{abstract}

\putsec{intro}{Introduction}

Network topology has a first order impact on the performance of general-purpose shared memory multiprocessor systems as it affects the latency and bandwidth of communication. 
Topology design must accommodate various conflicting demands in terms of cost 
(which may limit the number of links, the number of routers, the router degree) and 
performance (e.g., must achieve low latency on average, while achieving high 
bandwidth). Despite these complex trade-offs, the state-of-the-practice is to use 
topologies designed by human experts at the
chip/interposer~\cite{kite,butterdonut,doublebutterfly}. 
In contrast, unlike in the general-purpose system use-case, application-specific systems-on-chip (SoCs) have used 
automatic synthesis of networks-on-chip (NoCs) to exploit known application-specific communication patterns in topology design~\cite{lpbt,chatha2008automated,gangwar2020automatedsynthesis,murali2006designingnoc,bertozzi2005synthesis}. However, the challenges solved by these synthesis tools are not the same as those needed for general purpose systems where (1) communication is unknowable and likely random, and (2) the focus is on high performance and not necessarily resource minimization/power-efficiency (as is typical in the SoC context)\cite{lpbt,chatha2008automated,gangwar2020automatedsynthesis,murali2006designingnoc,bertozzi2005synthesis}. 
Naively applying these formulations in the general-purpose context has computational effort issues and yields networks that are worse (in both latency and throughput) compared to expert-designed networks, as we will show. Indeed, general purpose shared memory multiprocessor systems continue to use expert-designed networks predominantly ~\cite{sodani2016knightslanding,shim2023NCDE,bartolini2005recent}.

\figput[3in]{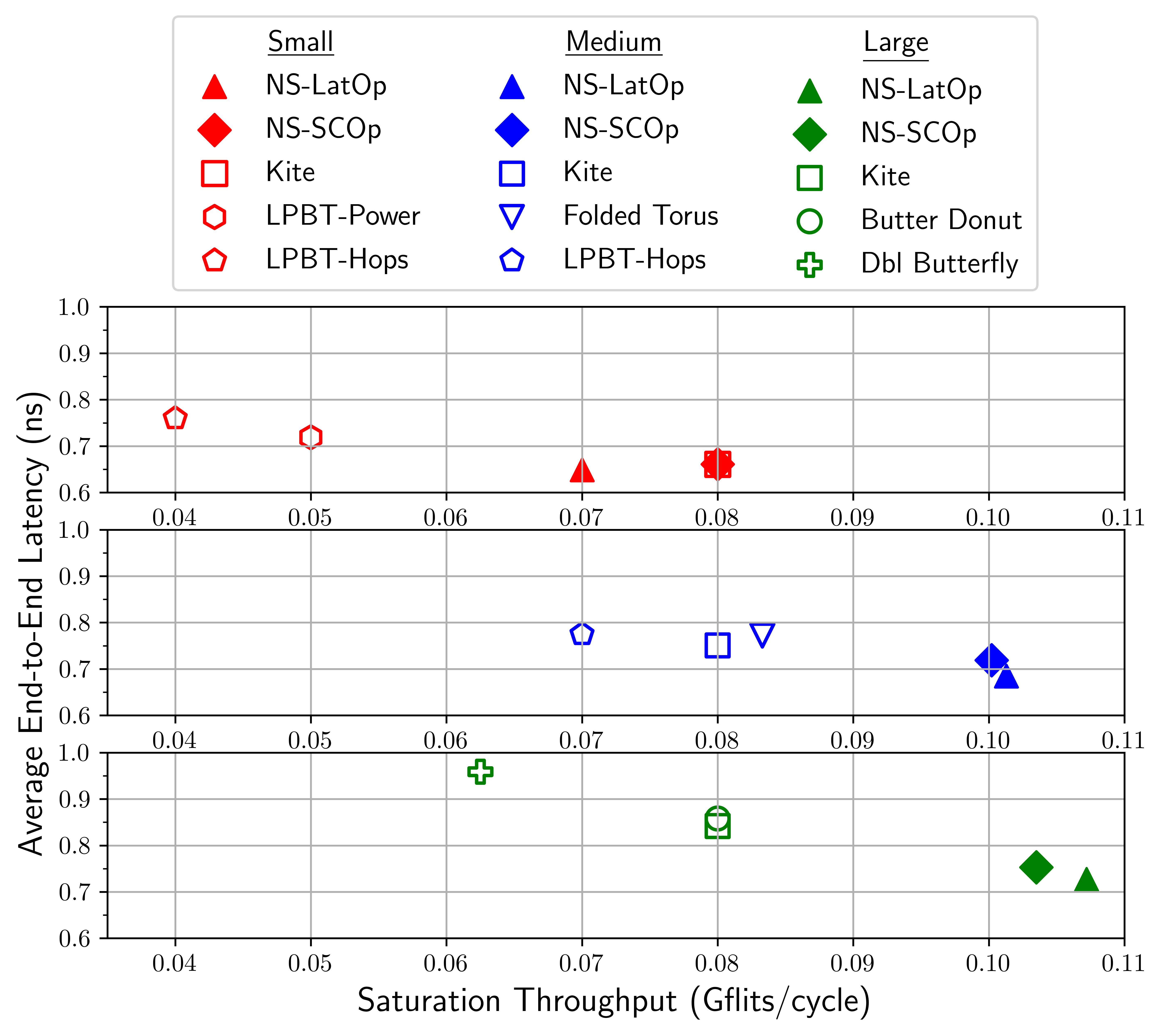}{}{Average packet latencies and saturation bandwidths of NoI topologies}

In this paper, we design and demonstrate an optimization-based framework -- \name 
 -- that automates topology discovery for such general purpose multicores. 
\name-generated network topologies focuses on generating topologies at the scale of networks on chips/interposers, which is the de facto scale for shared memory multiprocessor systems. 
\name's basic optimization framework uses the physical 
layout of routers, the link delay model, and the maximum acceptable link delay as 
inputs, and generates network topologies (complete with shortest path routing 
tables and deadlock-free VC assignment tables) that optimize one or more network 
performance metrics (e.g., average hop count, saturation throughput, or combined 
measures) as an objective function. 
To demonstrate \name's utility, we use it in the context of silicon interposer 
based systems with minimally active interposers which have been studied extensively 
recently~\cite{butterdonut,doublebutterfly,yin2018modularrouting,kite}. Because multiple network topologies have been 
proposed in this recent body of work (e.g., Butter Donut~\cite{butterdonut}, Double Butterfly~\cite{doublebutterfly}, Folded Torus, the Kite family~\cite{kite}), we are able to compare \name 
against these expert-designed networks. 
In addition, we also compare \name against prior optimization-based, NoC synthesis methods, namely \cite{lpbt}.

\figref{pareto} shows where each topology lies in terms of latency (Y-axis) and expected saturation throughput (X-axis) they achieve. Note that all topologies shown use the same number of routers and router radix, which makes this a cost-neutral comparison. The topologies are further grouped into ``small", ``medium" and ``large" categories based on their longest link lengths -- a taxonomy from cite{kite}.
Ideally, we want topologies that offer low latency and high throughput, which is the bottom right corner of the graph.
For now, we focus on the broad difference between legacy topologies and topologies generated by older automatic methods (with hollow markers) and \name-generated topologies (with solid markers and `NS' prefixed names); later in \secref{tops}, we offer a detailed analysis.  

In most cases, \name generated topologies offer strictly superior performance (lower latency {\em and} higher bandwidth) compared to expert-designed and previous optimization-based NoC synthesis (LPBT variants based on ~\cite{lpbt}) topologies. Further, \name is able to generate topologies that capture different latency/bandwidth trade-offs to generate topologies on the Pareto frontier.

One expert-designed topology -- Kite-Small (red hollow circle in \figref{pareto}) -- comes close (within 1\% of latency, same saturation throughput) to the \name generated optimal topology. These results conclusively make the case for \name by demonstrating that machine-generated topologies can transcend those designed by human experts.

We evaluate \name-generated topologies using the HeteroGarnet~\cite{kite} extension of gem5~\cite{gem5} simulator with synthetic and full-system traffic validate the performance expectation from the topology analysis. Area and power analysis using DSENT~\cite{sun2012dsent} show that the overheads are modest, and mostly rise from the aggressive topology design that maximally uses all available router ports according to the link-length budget.

In summary, we develop an optimization-driven topology generation framework called \textit{\name} that utilizes a novel MILP formulation to optimize for performance metrics of average hops and/or sparsest-cut, matching or outperforming both (1) past expert-designed networks, and (2) past topology generation frameworks with practical solver times. Though the \name-generated topologies are qualitatively more irregular (especially compared to expert-designed topologies), they pose no deployment challenges because
they are compatible with existing techniques for high-performance, deadlock-free routing (using table-based routing) and VC allocation. 
Evaluation by simulation reveals that \name-generated topologies for 4x5 interposer networks achieve up to 18-75\% higher saturation throughput for uniform random traffic than legacy networks. Further, the topologies reduce average packet delay of coherence and memory traffic by 10\% and overall execution time by 4\% on average across topology sizes for Parsec workloads compared to legacy networks. 
\putsec{background}{Background and Scope}
\putsubsec{interposer}{Interposer-Based Systems}
Due to a combination of factors including the slowing of Moore's law and the yield economics of larger chips, there has been increased interest in chiplet-based designs wherein smaller chiplets are integrated to form larger systems~\cite{hu2018diestacking,stow2017costeffective,lau2021chiplet,naffziger2021chiplet,isidor2023disintegratingmanycores}. We particularly focus on silicon interposer based systems in which the integration of multiple chiplets is achieved by stacking them on an interposer which offers basic communication and connectivity among the chiplets in the form of a network-on-interposer (NoI).
Because chiplets may internally have multiple cores, there may be a network-on-chip (NoC) within the chiplet, separate from the NoI. As explored in\cite{kite}, the NoC and NoI may be clocked at different frequency and are connected through clock domain crossings (CDCs).
There are a wide range of options for the interposer organizations ranging from passive designs with only metal layer links to fully active interposers with active logic. In this design space, Kannan \etal have argued for a {\em minimally active interposer} which does contain minimal logic (routers and repeaters) in the interposer, but the active area occupied is small enough that yields stay high. 
For such minimally active interposer systems, there have been recently proposed expert-designed network topologies~\cite{kite,butterdonut,doublebutterfly} that go beyond the common mesh topology for better performance. \name focuses on this class of networks. 

We only consider topologies that leverage the known best practices such as (1) the use of {\em mis-aligned} layouts~\cite{butterdonut} which allow cores at the chiplet boundaries to avoid interposer links, 
and (2) the use of {\em concentration}, wherein multiple cores/memory-controllers
are connected to a single router in the interposer. 
The above assumptions yields a simple regular placement of interposer routers in a 4x5 organization as shown in \figref{basic}(b). Each NoI router is connected to either four nearest cores in the chiplets (for the middle three columns of NoI routers) or two cores plus two memory controllers (for the NoI routers on the left-most and right-most columns). Given this layout, the specific network topology of the NoI is defined by the links between the interposer routers, which \name optimizes. Note that the above layout choice does not take away from \name's generality. Given any other layout and router radix, \name would similarly discover optimal network topologies for that layout as we show later with examples.

\figput[0.95\linewidth]{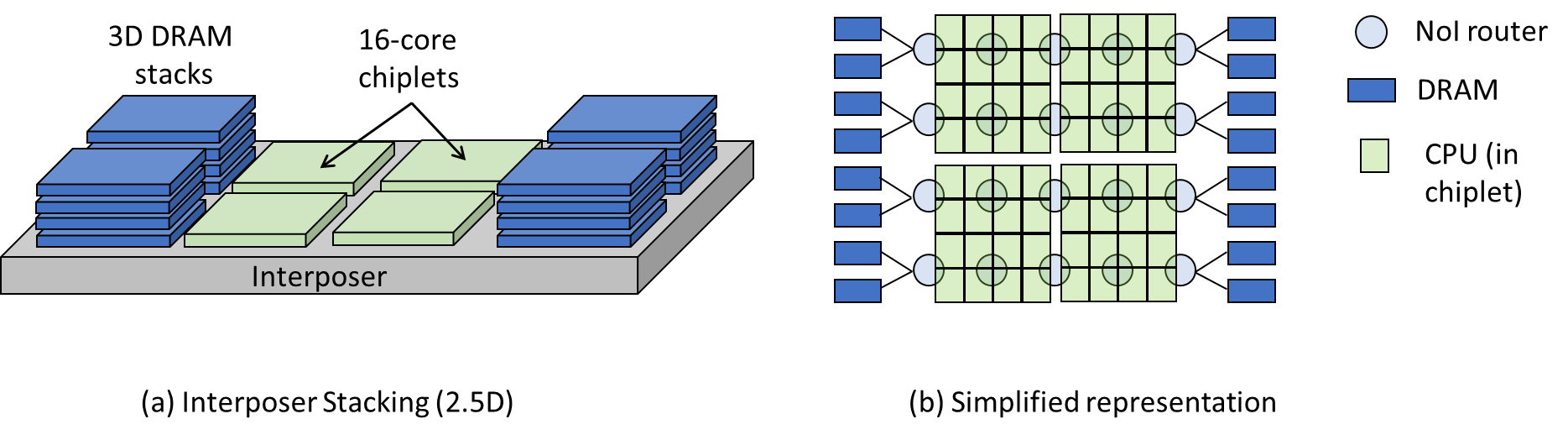}{basic2}{Example interposer-based system with 2.5D organization }

\putsubsec{scope}{Scope and Traffic Types}

Applications care about end-to-end packet latency and the throughput at which the network saturates. But reasoning about these performance metrics depends on the nature of traffic.
Because the split between coherence traffic and memory traffic can vary dynamically, in this paper, we focus broadly on improving average latency and throughput for uniform random traffic. Though our evaluations focus on homogeneous, all-to-all traffic, \mbox{\name} readily accepts other traffic patterns/models as input parameters for which it generates optimized topologies.
Specifically, we optimize for the average hop count and sparsest cut~\cite{jyoth2016measuringthroughput}\footnote{Though bisection bandwidth is a popular metric that attempts to capture the notion of a network-cut that acts as a throughput bottleneck, the {\em sparsest cut} is the more general definition of a cut-based throughput bottleneck as it is the tightest upper-bound on achievable throughput.} as proxies for latency and saturation throughput of the topology.

\putsubsec{latency}{Latency}

At low loads, where queueing delays are minimal, the end-to-end latency can be modeled as the average number of hops times the delay per hop -- dependent on technology parameters that affect wire and router delays. The observed average hops in practice may differ from that of the topology, particularly with non-minimal routing.
Although, with minimum-hop routing, the average hops of the topology directly determines that of the routed system.
The expected average hops of a topology can be calculated for a given traffic matrix. For general applicability, we use uniform, all-to-all so that no source and destination pairs are biased in optimization.
We optimize for the average hop count under uniform, all-to-all communication as an objective when generating topologies to reduce expected latency.

\putsubsec{throughput_bounds}{Throughput Bound(s)}

Network throughput bounds can be identified via bottleneck analysis, and the limiting bottleneck can be one of many possible causes. 
Some simple bottlenecks are due to limited injection/ejection ports. However, because these are bottlenecks that arise from local resource constraints at each router, it is straightforward to provision enough resources to avoid such throughput bottlenecks. 

The more interesting bottlenecks are (1) cut-based bottlenecks, and (2) link occupancy based bottlenecks which are properties of the topology. 
Cut-based bottlenecks effectively view the topology as a graph and identify graph-cuts that would be most loaded by network flows~\cite{jyoth2016measuringthroughput}. Bisection bandwidth is a widely used cut-based bottleneck, but the more general {\em sparsest cut} is the tightest possible cut-based bottleneck~\cite{jyoth2016measuringthroughput}. 
Link-occupancy based bottlenecks arise purely based on number of links/hops occupied by each flow. If the average distance traveled by each flow increases, more links are occupied, which in-turn increases channel load in aggregate. The best link-occupancy bounds are achieved by shortest-path routing because shortest paths minimize link occupancy. 
Finally, for a given routing algorithm, precise {\em maximum channel load} analysis can yield more accurate bounds on saturation throughput ~\cite{jyoth2016measuringthroughput}.

\putsubsec{routing}{Routing}

Table-based routing, which is a general form of routing, is already the routing method of choice for interposer networks~\cite{kite,butterdonut} and can be used to achieve minimal routing in \name generated topologies, independent of VC allocation (except expert-designed topologies).

The expert-design topologies proposed in recent works all utilize the same routing and deadlock avoidance scheme(s). Butter Donut, Double Butterfly, Folded Torus, and Kite topologies use shortest-path routing that adheres to a turn-based deadlock avoidance rule: no route may "double back" along the horizontal axis \cite{kite,butterdonut}.
We assume random selection of paths amongst the valid choices and call this routing ``no double back turns" or ``NDBT."
The MILP formulation in \cite{lpbt} that generated the LPBT topologies includes routing in the formulation
and we compare against the resultant routing function in our evaluations, labeling it ``LPBT" where applicable.

We employ a simple MILP formulation to minimize the maximum channel load bottleneck, or ``MCLB," and elaborate on the formulation in \secref{mclb-formulation}.
The concept of using MILP for routing is not new; however, the specific formulation used here has unique differences from previous works.
Overall, because we constrain the routing to shortest-paths, the set of possible paths is statically known and significantly shortcuts many of the previous formulations.
The MCLB formulation differs from \cite{capone2010routingschedulingandchannelassignment} which includes scheduling optimization, from algorithmic, mesh-specific \cite{vanchu2019lef}, and from \cite{pascual2018deadlockavoidanceforarbitrary} which includes VC allocation and deadlock avoidance in the routing algorithm.
Another approach \cite{kinsy2009deadlock} assumes the VC allocation for each flow (source, destination pair) is given and the optimization discovers routes amongst the given links under capacity constraints. In contrast, in MCLB, the set of all valid paths/routes is provided as input and the formulation simply selects which of the given paths results in the lowest maximum channel loads.

\putsubsec{deadlocks}{Deadlock Handling}

In the works that proposed the expert-designed topologies, the semi-regular networks avoid deadlocks through turn-based heuristic escape VC allocation by limiting routes that ``double back" along the horizontal axis and this scheme is utilized in our evaluations.
Deadlock handling for machine-generated topologies is not as straightforward as 
in some expert-designed topologies, even though the fundamental principles of deadlock freedom are the same. Dally and Seitz\cite{dally1988deadlock} showed that acyclic channel dependency graphs (CDG) are sufficient (but not necessary) to guarantee deadlock-free routing in wormhole networks
That original work has been expanded and extended in many ways~\cite{duato,ebrahimi2017ebda,lynse2006layeredrouting,domke2011deadlock,schwiebert2001deadlock,yin2018modularrouting,kinsy2009deadlock}, but minimally, it remains true at least one routing subfunction (e.g., on a subset of escape VCs) must remain cycle free to guarantee deadlock avoidance. Effectively, deadlock avoidance can at best pick two of shortest path routing, a limited number of virtual channels, or deadlock freedom. 
While \name cannot escape the fundamental nature of these constraints, \name optimizes the use of a limited number of escape VCs to achieve deadlock freedom for all the topologies we consider.
Other than simple turn-prevention schemes, all escape-path based deadlock-avoidance techniques and cycle-breaking deadlock-recovery schemes can be made to work with \name generated topologies.
\putsec{topgen}{Topology Generation}
\name's search space is constrained by (1) the layout of the routers, (2) the clocking/frequency limitations placed on link lengths, which indirectly constrains which router-pairs may be connected by links, (3) the need to achieve basic connectivity guarantees,
and (4) the need to respect hardware limitations (e.g. router radix).

Not all machine-generated topologies shown in \figref{pareto} are optimal, depending on the convergence of the optimization solver. However, with reasonable computational effort (e.g. minutes/hours), \name discovers high quality topologies that outperform expert-designed topologies. As such, our reported performance is a lower-bound on \name's capabilities. Note that our goal is to discover topologies that are better than expert-designed topologies. As long as the discovered topologies are better, it is less of a concern if complexity/tractability of the optimization problem prevents the discovery of the true optimal topology.

\putsubsec{mip}{Topology generation as an MIP optimization problem}

\tabref{mip_form} shows the definition of the key elements (i.e., objectives and constraints). We then use subsets of the objectives and constraints to generate designs with specific characteristics.
\name's formulation assumes that the router layout (which includes the number and physical placement of routers) and router radix are given. 
We define the total set of routers, $R$,
with integer labeled routers $0, 1,\ldots (r-1)$.

We use four sets of variables to track connectivity and minimum distance between routers.
The first set is a two-dimensional router-to-router connectivity map ($M$) of indicator variables in which the element at coordinate $(i,j)$ is set if the a link connects router $i$ to router $j$, and cleared otherwise \footnote{This can be further generalized to allow multiple connections between the same pair of nodes by changing the indicator variables to integers.}.
The second set of variables ($O$) represents the `one-hop' distance from any two routers $i$,$j$.
If two routers are not directly connected, their one-hop distance is infinity -- penalty in the distance constraint.
The third set of variables ($D$) is an integer, two dimensional matrix that tracks the minimum router-to-router hop distance.
An integer ($B$) represents the sparsest cut bandwidth and is defined through constraints.
All of the variables can be constrained and/or optimized in various combinations. Our focus is on optimizing for average hops and sparsest cut bandwidth with reasonable constraints.

\paragraph{Basic Network connectivity}
\name ensures two basic network properties. 
First, the number of links from each router cannot exceed the router radix, which can be counted by summing the elements of a row/column (outgoing/incoming radix constraint) in the connectivity map, as given as C2 in \tabref{mip_form}.
Second, all routers must be connected/reachable from all other routers, which can be restated as having a finite minimum hop distance to/from all other routers, stated as C8 in \tabref{mip_form} for any finite diameter. 

\paragraph{Link length}
One key feature of \name's topology search that enables reasonably fast discovery of solutions is that it constrains the length of network links to be no longer than an allowed maximum. 
The length constraint is driven by two reasons. First, there is the fundamental technological constraint that longer wires incur longer delays which may slow down the network. 
The second reason is that longer wires enlarges the search space for the optimization problem as it increases the number of possible connections between routers. 
Given that the wire-delays limit the use of long links anyway, \name is designed to operate within such user-chosen limits to avoid unnecessary slowdown. 

We refer to link length limits using the same 
nomenclature developed in Kite\cite{kite} as 
illustrated in \figref{linklength}, which names links based on the hops they span in the X and Y dimensions. Specifically, 
we also use the classification developed in Kite
to refer to networks with a limit of (1,1) links as 
{\em small}, (2,0) links as {\em medium},  and (2,1) 
links as {\em large}. In order to constrain the MIP formulation to keep all links within the length limitations, we populate a set of valid links, $L$, and constrain the connectivity map as given by C3 of \tabref{mip_form}.

\figput[0.6\linewidth]{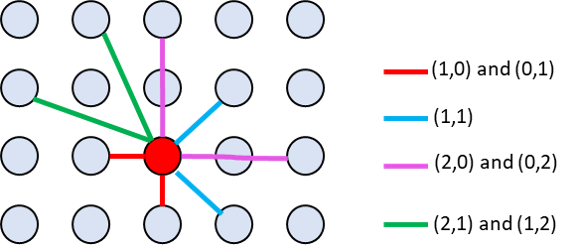}{}{Single-hop reach with varying link length}

\begin{table*}[htbp]

\caption{\label{tab:mip_form} Constraints + Objectives for \name Topology Generation}
\vspace{-1.0em}

\begin{tabular}{|c|>{\centering}m{0.375\linewidth}|p{0.375\linewidth}|c|c|}

\hline 
\bf Label & \bf Objective/Constraint & \bf Description & \bf LatOp & \bf SCOp \\
\hline \hline 
O1 & $D_{total}=\sum_{s \in R}\sum_{d \in R}D(s,d)$&Latency objective function (minimization) &$\checkmark$&\tabularnewline
\hline

O2 & $\begin{array}{c}
B_{total}=\min_{(U,V)} B(U,V)\end{array}$
& Sparsest cut objective function (maximization)
&&$\checkmark$\\

\hline 

\hline 
C1 & $\begin{array}{c}
\forall i \in R,\quad M(i,i)=0, D(i,i)=0
\end{array}$& Ignore self-adjacency &
$\checkmark$&$\checkmark$\\
\hline 

C2 & $\begin{array}{c}
\forall i \in R,\quad \sum_{j \in R} M(i,j) \leq radix\\
\quad \sum_{j \in R} M(j,i) \leq radix\end{array}$ & Out/In router radix&$\checkmark$&$\checkmark$ \tabularnewline
\hline

C3 & $\begin{array}{c}
\forall i,j \in R,\quad \text{if } (i,j) \notin L \text{ then } M(i,j)=0 \end{array}$
&Link length &$\checkmark$&$\checkmark$\tabularnewline
\hline 

C4 & $\begin{array}{c}
\forall i,j \in R,\quad \text{if } M(i,j)==1 \text{ then } O(i,j)=1 \\
\text{ else, } O(i,j)=\infty \end{array}$
& Enforce single-hop path length &$\checkmark$&
\tabularnewline
\hline 

C5 & $\begin{array}{c}
\forall i,j \in R,\quad D(i,j)=min(D(i,k) + O(k,j)) \\ \forall k \neq j, i \neq j \end{array}$
& Multi-hop shortest path&$\checkmark$&
\tabularnewline
\hline

\hline 
C6 & $\begin{array}{c}
\forall U, V \subset R,\quad B(U,V)= \sum_{i \in U} \sum_{j \in V} \frac{M(i,j)}{|U||V|} \end{array}$ &
Exhaustive sparsest cut bandwidth computation &&$\checkmark$ \tabularnewline
\hline 

C7 & $\begin{array}{c}
\forall U, V \subset R,\quad B(U,V) \geq {min\_bandwidth}
\end{array}$&
Minimum sparsest cut bandwidth&&$\checkmark$
\tabularnewline

\hline 
C8 & $\begin{array}{c}
\forall i \in R,\quad \sum_{j \in R} D(i,j) \leq diameter\end{array}$& (Optional) Bounding network diameter&$\checkmark$&$\checkmark$\tabularnewline
\hline 
C9 & $\begin{array}{c}
\forall i,j \in R,\quad M(i,j)=M(j,i)\\
\end{array}$& (Optional) Link symmetry &&\tabularnewline
\hline

\end{tabular}
\end{table*}

\paragraph{Link Symmetry}
\name can trivially support symmetric links wherein  a link from router $R_A$ to router $R_B$ is always implicitly paired with a link in the opposite direction from $R_B$ to $R_A$. (See constraint C9.)
Alternately, if one were to eliminate the constraint, 
\name can also consider asymmetric links, wherein the outgoing half of the full duplex link can connect to a different router than the incoming half. For example,  the Kautz-graph topology used in the SiCortex machines~\cite{stewart2006kautz}) used such asymmetric links. We found that asymmetric links yields a modest gain in throughput (3\%), and thus use asymmetric links in our results. Note, if one were interested only in designs with symmetric links, \name would still offer improvements over existing designs.

\paragraph{Hop Distance}

We constrain a two dimensional integer matrix, $D(i,j)$, to represent all source to destination shortest paths using a triangle inequality~\cite{clr} constraint.
Specifically, each element $d_{i,j}$ is constrained to equal the minimum value of the set of all distance of paths through a third router $k$. This constraint is given as C5 in \tabref{mip_form}.
The path distance from $i$ to $j$ through $k$ is defined as the sum of hop distance $i$ to $k$ and the one-hop distance $k$ to $j$. The special case when $i == k$ represents the one-hop distance between $i/k$ and $j$. The special case $k == j$ is the same path $i$ to $k/j$ and is not considered in the set of routes through k because creates a self referencing constraint. The case $i == j$ is omitted as C1 defines the value explicitly.

To formulate the triangle inequality, we define an intermediary two dimensional integer variable, $O(i,j)$, to represent the one-hop distances from all source to destination nodes and equals connectivity matrix for connected routers. If two routers are directly connected then their one-hop distance is equal to one; otherwise, the one-hop distance equals ``infinity" (a large number in practice).
Gurobi provides a high-level constraint for if-then conditions as a convenience for users like C4 in \tabref{mip_form} for brevity but is modeled through indicator variables in actuality.
No other MIP-based NoC synthesis work defines an average hops metric constrained by a triangle inequality~\cite{lpbt,chatha2008automated,huang2012applicationspecific,zhong2011applicationspecific,mukherjee2016ilpfloorplan}.
Bounding the diameter, C8, is optional but helps reduce time to find the first solution.

\paragraph{Sparsest Cut Bandwidth}

We constrain a scalar integer variable, $B$, to represent the sparsest cut bandwidth of the topology. This variable is constrained to the minimum of all sparsest cut bandwidths as stated in C6 of \tabref{mip_form}. We constrain against all possible combinations of partitions, $U$,$V$ of the routers. For the twenty router configuration, that is $20!$ combinations. Although combinations grow quickly, we found calculating the sparsest cut for the twenty router configurations to be feasible in reasonable time frames as the explosion in constraints also reduces the solution space.

For each sparsest cut, the bandwidth $B(U,V)$ is calculated as the number of direct router connections between all routers from $U$ to $V$ divided by the product magnitudes $|U|$ and $|V|$, which scales each bandwidth for uniform, coherence traffic.
After constraining the sparsest cut bandwidth variable, we allow setting the minimum sparsest cut bandwidth by requiring that all cuts satisfy the minimum value, as given as C7 in \tabref{mip_form}. For asymmetric/unidirectional links, we consider the minimum bandwidth among the two directions because that is the true bottleneck. No other MIP-based NoC synthesis work defines a sparsest cut bandwidth metric~\cite{lpbt,chatha2008automated,huang2012applicationspecific,zhong2011applicationspecific,mukherjee2016ilpfloorplan}.

\paragraph{Total Hop Count Objective}

We reduce the expected latency for network topologies by minimizing the total - equivalently, average - hop count of the topology as the sum of all the source-destination distance variables.

\figput[0.6\linewidth]{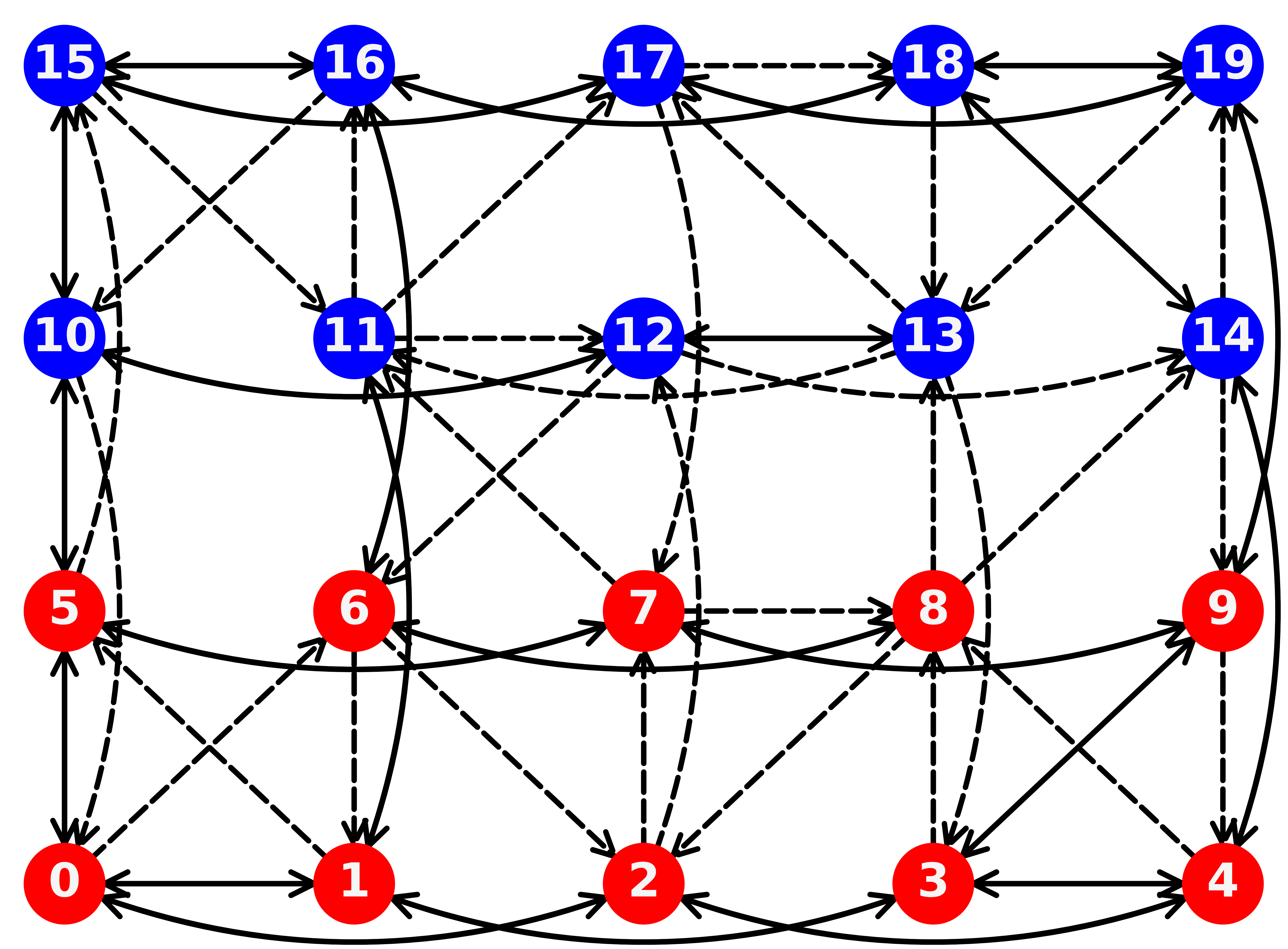}{}{Latency-optimized \name medium topology}

\paragraph{Sparsest Cut Bandwidth Objective}

We increase the expected saturation throughput for network topologies by maximizing the sparsest cut bandwidth of the topology, as defined by C7.
Packets cross bottlenecks/cuts at loads proportional to the source/destination set magnitudes. The maximally loaded cut may be any selection of source/destination sets: maximum channel load is not limited to the bisection case. Furthermore, there are no patterns/subsets to the discovery of sparsest cut.
As such, the exhaustive approach to identify the minimum bandwidth cut helps avoid such overstatement. Later in \secref{results} we see that the saturation bandwidth more or less follows the true sparsest cut bandwidth (as long as no other bottlenecks exist).

\putsubsec{tops}{\name-Generated Topologies}

\name generates topologies with specific optimization objectives by selecting the constraints and objectives from \tabref{mip_form}. Specifically, we focus on two variants: latency-optimized ({\em LatOp}), and sparsest cut bandwidth-optimized ({\em SCOp}).
We use Gurobi v9.5.1 as our solver to find topologies \mbox{\cite{gurobi}}. Specifically, we use the Gurobi C++ library. We run each Gurobi optimization problem on 2.8GHz AMD Opteron Processor 6320s with 256GB of memory and a maximum of 32 threads.

\figref{ns_m_latop} illustrates an example latency-optimized topology generated by \name for the medium link-length constraint.
The topology is colored to indicate an example sparsest cut (red nodes in one partition and blue nodes in the other) and for this example, the sparsest cut is indeed a bisection.
One visual remark is that because the topologies can have asymmetric links, they may appear ``busier" than symmetric networks. But the underlying hardware resource usage is the exact same as that of topologies with symmetric links. To improve visibility of \figref{ns_m_latop}, bidirectional and unidirectional links are plotted in solid and dashed lines, respectively.

Quantitatively, \tabref{metrics} shows the key topology metrics for \name-generated topologies along with recent expert-designed topologies. In addition to expert-designed topologies, it includes the LPBT topologies, generated by ~\cite{lpbt}.
The average hops is calculated as the simple average of all source-destination pairs -- not including routing a node to itself -- without weighting for specific traffic patterns.
The bisection bandwidth is provided instead of the sparsest cut for ease of comparison with other works as they report only bisection bandwidth.
Uniformly, \name manages to outperform expert-designed topologies via its optimization approach. In the medium and large configurations, \name outperforms on both bisection bandwidth (by 50\% and 75\%, respectively) and latency (8\% and 13.5\%, respectively) compared to the best metrics of any other topology. Folded Torus has a bisection bandwidth of 10 and all other large topologies considered have bisection bandwidths of 8. Kite-Medium and Kite-Large topologies have the lowest average hop values of the human-generated topologies at 2.32 and 2.27, respectively.

In the small configuration, one expert-designed topology (Kite-Small) comes close to the optimal topology discovered by \name. Compared to Kite-Small, our latency optimized topology improves the latency by 1.7\% at the cost of a 13.5\% drop in bisection bandwidth. Our sparsest cut topology is actually equal to Kite Small, indicating that Kite Small is (unknowingly) the optimal topology given the parameters.

\name's link length distributions are not skewed towards an increased number of the longer links. In fact Kite-large, has a higher number of long (2,1) links compared to \name.
For small topologies, all machine-generated topologies, \name and LPBT included, utilize the (1,0) links more than Kite. 
In the medium category, \name and Kite have similar proportions of links but disparate average hop counts, demonstrating the importance of which connections are made.

We include results for a 30 router configuration in \tabref{metrics}, where the design rules for previous topologies are logically extended. Without iteratively designing heuristic architectures, \name systematically produces topologies better in every metric for each size in a reasonable amount of time, as will be seen. Although the larger design space implies increased solver effort, the room for performance improvement increases as well.

Finally, from data not presented, we make two observations on the topologies generated by \name. First, we observe that the reduction in average hops is achieved by shifting the whole latency distribution downward; and not the net effect of some increases and some decreases. 
Second, we observe that forcing symmetric links leads to a loss of under 3\% in latency compared to the topologies with asymmetric links and no loss in bandwidth. As such, there is value in using \name even if one wanted to avoid asymmetric links.

% \usepackage{multirow}
\tabput{metrics}{
\small
\begin{tabular}{|c|ccccc|}
\hline
\textbf{\begin{tabular}[c]{@{}c@{}}\#\\ Routers\end{tabular}} & \multicolumn{1}{c|}{\textbf{Topology}} & \multicolumn{1}{c|}{\textbf{\begin{tabular}[c]{@{}c@{}}\#\\ Links\end{tabular}}} & \multicolumn{1}{c|}{\textbf{Diam.}} & \multicolumn{1}{c|}{\textbf{\begin{tabular}[c]{@{}c@{}}Avg.\\ Hops\end{tabular}}} & \textbf{\begin{tabular}[c]{@{}c@{}}Bi.\\ BW\end{tabular}} \\ \hline \hline 

                                                              & \cellcolor[HTML]{DAE8FC}Kite           & \cellcolor[HTML]{DAE8FC}38                                                       & \cellcolor[HTML]{DAE8FC}4           & \cellcolor[HTML]{DAE8FC}2.38                                                      & \cellcolor[HTML]{DAE8FC}8$^\filledstar$                                 \\
                                                              & LPBT-Power                             & 33                                                                               & 5                                   & 2.59                                                                              & 4                                                         \\
                                                              & \cellcolor[HTML]{DAE8FC}LPBT-Hops      & \cellcolor[HTML]{DAE8FC}34                                                       & \cellcolor[HTML]{DAE8FC}6           & \cellcolor[HTML]{DAE8FC}2.74                                                      & \cellcolor[HTML]{DAE8FC}4                                 \\
                                                              & NS-LatOp$^\filledstar$                               & 37                                                                               & 4                                   & 2.34$^\filledstar$                                                                              & 7                                     \\
                                                              & \cellcolor[HTML]{DAE8FC}NS-SCOp$^\filledstar$        & \cellcolor[HTML]{DAE8FC}37                                                       & \cellcolor[HTML]{DAE8FC}4           & \cellcolor[HTML]{DAE8FC}2.38                                                      & \cellcolor[HTML]{DAE8FC}8$^\filledstar$                     \\ \cline{2-6} 
                                                              & Folded Torus                           & 40                                                                               & 4                                   & 2.32                                                                              & 10                                                        \\
                                                              & \cellcolor[HTML]{DAE8FC}Kite           & \cellcolor[HTML]{DAE8FC}40                                                       & \cellcolor[HTML]{DAE8FC}4           & \cellcolor[HTML]{DAE8FC}2.25                                                      & \cellcolor[HTML]{DAE8FC}8$^\dagger$                                 \\
                                                              & LPBT-Hops                              & 38                                                                               & 4                                   & 2.33                                                                              & 7                                                         \\
                                                              & \cellcolor[HTML]{DAE8FC}NS-LatOp       & \cellcolor[HTML]{DAE8FC}40                                                       & \cellcolor[HTML]{DAE8FC}4           & \cellcolor[HTML]{DAE8FC}2.06                                                      & \cellcolor[HTML]{DAE8FC}10                                \\
                                                              & NS-SCOp                                & 40                                                                               & 4                                   & 2.16                                                                              & 11$^\filledstar$                                                        \\ \cline{2-6} 
                                                              & Butter Donut                           & 36                                                                               & 4                                   & 2.32                                                                              & 8$^\dagger$                                                         \\
                                                              & \cellcolor[HTML]{DAE8FC}Dbl. Butterfly & \cellcolor[HTML]{DAE8FC}32                                                       & \cellcolor[HTML]{DAE8FC}4           & \cellcolor[HTML]{DAE8FC}2.59                                                      & \cellcolor[HTML]{DAE8FC}8                                 \\
                                                              & Kite                                   & 36                                                                               & 5$^\dagger$                                   & 2.27                                                                              & 8$^\dagger$                                                         \\
                                                              & \cellcolor[HTML]{DAE8FC}NS-LatOp       & \cellcolor[HTML]{DAE8FC}40                                                       & \cellcolor[HTML]{DAE8FC}3           & \cellcolor[HTML]{DAE8FC}1.96                                                      & \cellcolor[HTML]{DAE8FC}13                                \\
\multirow{-15}{*}{20}                                         & NS-SCOp                                & 40                                                                               & 4                                   & 2.03                                                                              & 14                                                        \\ \hline \hline 

                                                              & \cellcolor[HTML]{DAE8FC}Kite           & \cellcolor[HTML]{DAE8FC}58                                                       & \cellcolor[HTML]{DAE8FC}5           & \cellcolor[HTML]{DAE8FC}2.91                                                      & \cellcolor[HTML]{DAE8FC}10                                \\
                                                              & NS-LatOp                               & 58                                                                               & 5                                   & 2.80                                                                              & 8                                                         \\ \cline{2-6} \cline{2-6} 
                                                              & \cellcolor[HTML]{DAE8FC}Folded Torus   & \cellcolor[HTML]{DAE8FC}60                                                       & \cellcolor[HTML]{DAE8FC}5           & \cellcolor[HTML]{DAE8FC}2.79                                                      & \cellcolor[HTML]{DAE8FC}10                                \\
                                                              & Kite                                   & 60                                                                               & 5                                   & 2.66                                                                              & 10                                                        \\
                                                              & \cellcolor[HTML]{DAE8FC}NS-LatOp       & \cellcolor[HTML]{DAE8FC}59                                                       & \cellcolor[HTML]{DAE8FC}5           & \cellcolor[HTML]{DAE8FC}2.47                                                      & \cellcolor[HTML]{DAE8FC}11                                \\ \cline{2-6} 
                                                              & Butter Donut                           & 44                                                                               & 10                                  & 3.71                                                                              & 8                                                         \\
                                                              & \cellcolor[HTML]{DAE8FC}Dbl. Butterfly & \cellcolor[HTML]{DAE8FC}48                                                       & \cellcolor[HTML]{DAE8FC}5           & \cellcolor[HTML]{DAE8FC}2.90                                                      & \cellcolor[HTML]{DAE8FC}8                                 \\
                                                              & Kite                                   & 56                                                                               & 5                                   & 2.69                                                                              & 10                                                        \\
\multirow{-9}{*}{30}                                          & \cellcolor[HTML]{DAE8FC}NS-LatOp       & \cellcolor[HTML]{DAE8FC}60                                                       & \cellcolor[HTML]{DAE8FC}4           & \cellcolor[HTML]{DAE8FC}2.32                                                      & \cellcolor[HTML]{DAE8FC}14                                \\ \hline

\multicolumn{6}{c}{$\filledstar$ known optimal topology/metric}\\
\multicolumn{6}{c}{$\dagger$ differ from values given by \cite{kite,butterdonut}}
\end{tabular}

}{Topology Metrics}
% end new table

\putsubsec{comp-effort}{\name's Computational Effort}

\figput{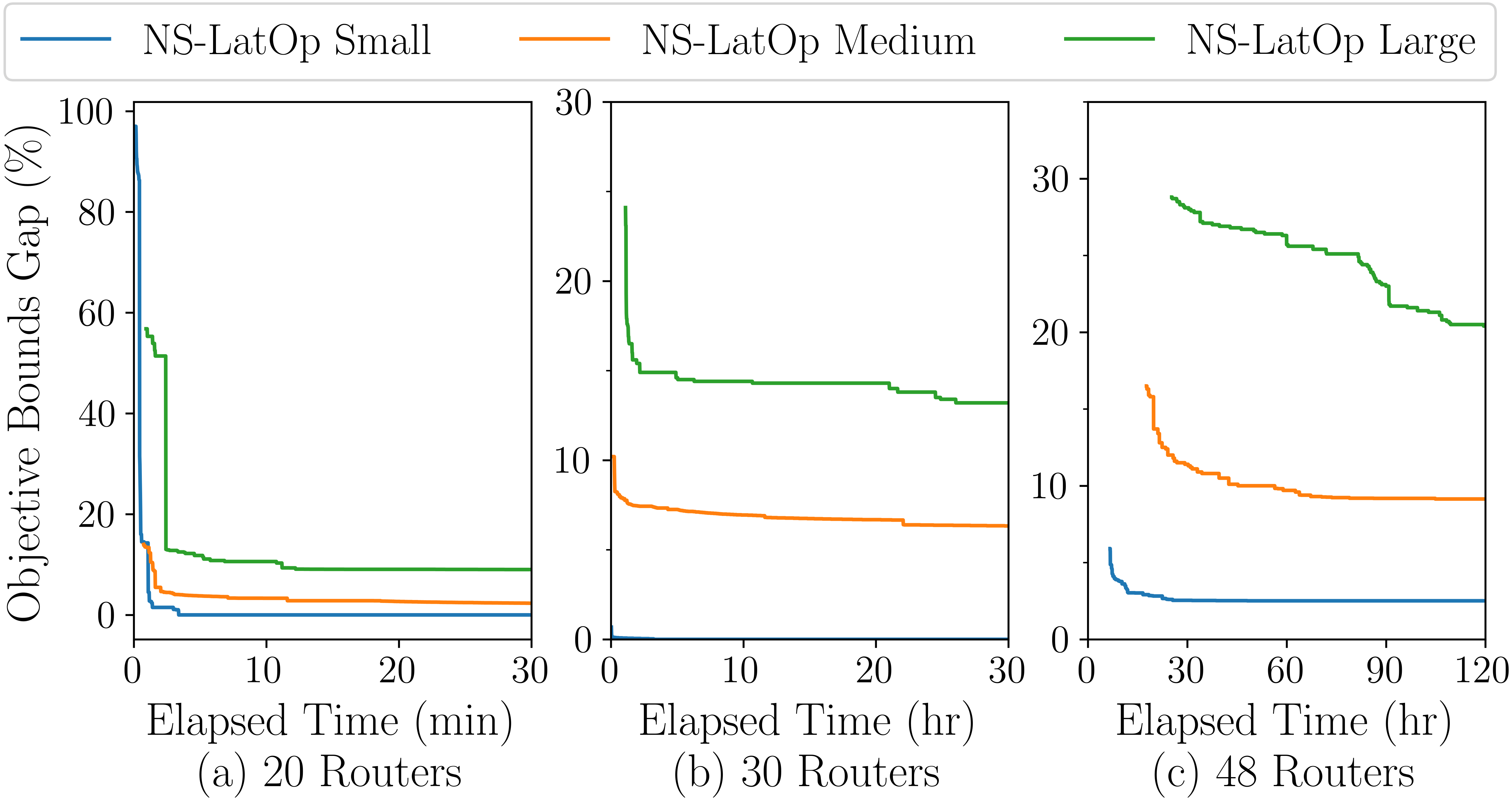}{}{\name solver progress over time}

Because \name relies on MIP optimization and because MIP, in general, is known to be NP-hard~\cite{hartmanis1982miphard}, there are no guarantees on the computational feasibility for truly optimal solutions for \name's formulations.
However, we rely on the fact that heuristic branch-and-bound solvers are effective in practice for many specific optimization problems to measure \name's optimization time.
 
MIP solvers typically track the ``objective bounds gap" metric which is the maximum gap possible between the currently known solution and any possible optimal solution. The metric is a measure of uncertainty and not of optimality; the current known solution may in fact be optimal but the solver may not be certain. 
\figref{progress}(a) shows how the objective bounds gap (Y-axis) narrows with solver time (X-axis) for each of the {\em small, medium,} and {\em large} configurations our twenty router NoI using \name's latency-optimized (LatOp) MIP formulation. 

We make two observations from the results.
First, the smaller the link length limit, the faster the convergence to a good solution. 
The {\em small } configuration converges to the optimal solution in under 5 minutes. The medium takes longer, but comes close to optimal (within 3\%) in about 13 minutes.
The large configuration only shrinks the objective bounds gap to about 9\% in 30 minutes. 

Second, even in cases where the objective bounds gap plateaus without converging to near zero in a reasonable amount of time, the resulting topology can be significantly better than any expert-designed topology. Indeed, \name's latency optimized MIP search for large topologies only reached approximately 7\% objective bounds gap; but it yields a design that has 13.5\% lower latency and 63\% higher throughput than any expert-designed large network. 

\figref{progress}(b) shows that scaling up the network to 30 nodes (with 6x5 layout) for the topologies given in \tabref{metrics} yields similar relative trends but with an increase in absolute times (hours instead of minutes). 
A similar trend is seen in \figref{progress}(c) with a 48 node, 8x6 layout which seems to reach a saturation point within two days. The 48 node layouts were superior to any heuristics as will be shown in \secref{scalability-results}.
The 30 node configuration also outperformed heuristic topologies in simulation but results were omitted due to  space constraints.
In addition, we performed other scalability studies to increase the number of routers and the number of ports per router. 
Surprisingly, increasing the router radix decreases convergence time and results in faster solutions but these results are not shown due to lack of space. 

In comparison, the previous MILP NoC synthesis framework (LPBT \cite{lpbt}) was significantly slower. The native design, which optimizes for NoC power (LPBT-Power) generated the first candidate small topology after {\em 20 days} of compute effort for the small link lengths with twenty routers. We modified the formulation to minimize an intermediate "latency" variable in the formulation; however, only the small and medium configurations produced  a (non-optimal) topology (LPBT-Hops) after {\em 20 days}.
The formulation for latency in LPBT is a sum of all possible hops for all links between a source and destination while the hop count constraint in \name utilizes a triangle inequality. Therefore, the latency formulation in LPBT and \name are distinctly different, due to which \name produces higher quality solutions in a shorter (and feasible) amount of time.

\putsubsec{mclb-formulation}{MIP Routing}

We also utilize MIP to determine path selection to best utilize the irregular \name topologies.
Our routing contribution, MCLB, essentially solves optimal path selection among available shortest paths: the formulation is provided a flat list, $P$, of all possible/desired paths between all sources and destinations and returns a flat list of chosen paths -- one per flow -- between all sources and destinations such that the maximum load across any link is minimized. 
The set of all paths of $P$ between a source, $s$, and destination, $d$, is calculated statically from the topology using the Floyd-Warshall algorithm and organized into a path set $P[s][d]$. This set is the only necessary input to the MCLB formulation.

A four-dimensional matrix, $flow\_load$, describes the load on a link from each flow.
For simplicity, the selection of paths is constrained to single-path and uniform demand which implies $flow\_load[s][d][i][j]$ can be represented as a binary one-hot of whether a path, $s$ to $d$, traverses link between routers $i$ and $j$.
An integer two-dimensional matrix, $cload$, represents the cumulative load on a link.
Intermediate indicator arrays, $link\_used$ and $path\_used$, for each flow and utilized to describe the selection of links and paths, respectively. These variables constrain and define the flow load on a link, $flow\_load$.

The 20 router NoI configuration used for synthetic evaluations used true shortest-path routing and takes under 5 minutes to complete.
Routing the 84 router, full-system configuration takes 5-10 minutes on the hardware described in \secref{tops} and the majority of runtime is setting up the formulation.

\paragraph{Objective}

MCLB optimizes for the minimum maximum channel load as given in O1 of \tabref{mclb_form} and is a common \textit{minmax} problem. In implementation, an intermediary variable is constrained as greater than or equal to all channel loads and minimized in the objective function.

\paragraph{Load Constraints}

The channel load over a link $(i,j)$ is simply the sum of all load from all flows as described by C1 \tabref{mclb_form}. For this simple use case, the flow is uniform, all-to-all but it can be extended to specific traffic patterns.

\paragraph{Path Selection}

For each source and destination path, a intermediate binary array, $link\_used$, of length equal to number of hops of a path (i.e. channels), is used to map the four-dimensional $flow\_load$ matrix to a single array representing this path.
If a path is selected as the route between a source and destination then is sets all $link\_used$ along the path which maps to set high the corresponding $flow\_load$ element.
Whether a path was selected/used is defined as an AND condition between all $link\_used$ indicator variables for all links along the path. Remark, $link\_used$ values \textit{not} on the path are unconstrained. Particularly, on links that do not affect the maximum channel load, the status of the $link\_used$ variable is inconsequential.

\paragraph{Single-Path Routing}
Specifying that only one path between a source and destination is asserted as special ordered set amongst the indicator variables of which path was selected, or equivalently as given by C4 in \tabref{mclb_form}, as simply setting the summation of all (binary) $path\_used$ variables to 1. This criteria can be modified to accommodate fractional or multi-path routing.

\begin{table}[]
\caption{\label{tab:mclb_form} Constraints + Objectives for MCLB Routing}
\vspace{-1em}
\begin{tabular}{|l|l|l|}
\hline
Label & Objective/Constraint & Description \\ \hline
\hline
O1 & $C_{total} = \max_{(i,j)} cload[i][j] $ & \makecell[l]{Maximum channel \\ load (minimization) }\\ \hline
\hline
C1 & \makecell[l]{$cload[i][j] =$ \\ $\sum_{s} \sum_{d} flow\_load[s][d][i][j] \forall i,j \in R$} & Channel load \\ \hline
C2 & \makecell[l]{$link\_used[i] = flow\_load[s][d][n_{i}][n_{j}]$ \\ $\forall n_{i} \text{ of k length path }  \{n_0,n_1,...n_{k-1}\}$ \\ $0 \leq i \leq k - 2, \forall s,d \in R$ } & \makecell[l]{Link utilization \\ along path } \\ \hline
C3 & \makecell[l]{$path\_used[p] = \prod_{i} link\_used[i] $ \\ $\forall p \in P[i][j]$ } & Path used criteria \\ \hline
C4 & \makecell[l]{$1 = \sum_{p \in P[s][d]} path\_used[p] $ \\ $\forall s,d \in R$ } & \makecell[l]{Single path \\ routable/connected } \\ \hline
\end{tabular}

\end{table}
\putsec{method}{Evaluation Methodology}

\tabput{param}{
% \small
\vspace{-1.0em}
\begin{tabular}{|r|p{2.6in}|}
% \multicolumn{2}{c}{\bf Baseline CPU}\\
\hline
Core & 4 chiplets, 64x OoO core,  3.8GHz, \\
 & 32kB L1D/I, 2MB L2 \\ 
Memory & 16x 2GB DDR4, 3.8GHz\\
\hline
Network & 3.8GHz 4x 4x4 mesh NoC, 4x5 NoI router topology\\
& 8B link width, 2/2 cycle router/CDC latencies \\
& MCLB \& LPBT: 6 escape VCs \\
& NDBT: 2 escape VCs \\
%  & 2 cycle router latency\\
Protocol & MESI Two Level, 10 total VCs\\
\hline
Kernel/OS & Linux 4.19.83, Ubuntu 18.04.2 LTS\\
\hline
 
\end{tabular}
}
{Full System Parameters}

We use gem5~\cite{gem5} v22 and the HeteroGarnet~\cite{kite} extension to evaluate the effect of the interposer technologies on network communication for synthetic and real workloads. 
We isolate the performance impact of the interposer through synthetic traffic and a simplified 20 CPU, 8/20 directory system directly attached to the proposed NoI topologies at a 1:1 ratio.
For the synthetic traffic evaluation, we measure the average packet latency across a sweep of injection rates of uniform random traffic using the Garnet Synthetic Traffic Generator and Garnet Standalone protocol \cite{gem5}. 
Control (8B) and data packets (72B) are injected with equal likelihood. We allocate 6 total VCs for all routing schemes, allowing each scheme to utilize more/less escape virtual channels as required.

In evaluating real workloads, we simulate 64 out-of-order CPUs
in a configuration similar to \figref{basic}, 
with parameters as given in \tabref{param}.
For the real workloads, we measure the execution time relative to mesh to determine speedups.
We simulate all PARSEC benchmarks (except $vips$ with baseline errors) to represent a range of network loads \cite{bienia2008parsec}. The benchmarks were simulated by limiting simulation on all CPUs to a maximum of 100 million instructions of a region of interest with 100 million instructions of warmup.

For both measurements, the NoI is clocked at the fastest clock speed allowable for its longest link -- as described in \cite{kite} -- where small, medium, and large topologies are clocked at 3.6GHz, 3.0GHz, and 2.7GHz, respectively.
We verify timing constraints for the topologies are met with DSENT \cite{sun2012dsent}.
We omit comparisons with networks that have repeatedly been shown to have poor metrics (e.g., mesh, concentrated mesh), and (2) have aligned router placement, as they are known to be worse than those with ``misaligned" router placement -- as confirmed in our own experiments.

\putsubsec{routevc}{Routing and VC Allocation}

We apply the routing and VC allocation schemes of prior works. 
Specifically, we use shortest path routing with no double-back turns (NDBT), to the expert-designed topologies and LPBT's internally defined routing for the LPBT topologies.
The LPBT-generated topologies do not specify; so we apply our allocation scheme to achieve deadlock freedom\cite{lpbt}.
We make one modification to the criteria for path selection in the full-system configuration motivated by the high latency incurred CDC crossing between the NoC and NoI. The set of valid paths is constrained to those that do not double-back from NoC to NoI or vice-versa, minimizing it to the required crossings.

Because such shortest path routing may induce cycles in the channel dependency graph (CDG) of the network, we partition the set of shortest paths such that the paths in each partition do not induce cycles in the CDG. Each such partition is then assigned to a virtual channel (VC). 
The optimal allocation of source-destination paths to virtual channels is known to be NP-Hard \cite{domke2011deadlock}. We apply the DFSSSP algorithm as given in \cite{domke2011deadlock} which iteratively removes shortest paths that induce cycles to a new VC.
We found that simple, random selection of the cycle-forming back edge to be removed at each iteration gave sufficiently low required virtual channels.
This method yields deadlock-free VC assignments using only 4 VCs for all 20-router configurations. Note, the outlier is an expert-design topology, Folded Torus, that bounds the minimum escape virtual channels to 4.
The virtual channels were load balanced using path-length weighted VC occupancy as the metric (e.g. a path traversing three links has a weight of three). 
\putsec{results}{Experimental Results}
In addition to the analytical results presented in earlier sections, 
we present an experimental evaluation of \name-generated topologies with (1) synthetic traffic and (2) real application workloads. 
We breakdown the effect of routing on these results and also present the area/power analysis of \name generated topologies.

\putsubsec{synthetic-results}{Synthetic Traffic}
\figref{main-flat_coh} illustrates the achieved latency (Y-axis) and throughput (X-axis) achieved
as we vary the injection rate of synthetic uniform random traffic, for the small, medium, and large topologies. 
One common trend is the expected pattern of low latency at low loads, with gradual increase as offered injection rate increases, with a sudden latency degradation at network saturation. 
Though the three classes of topologies operate at different clock speeds, the saturation throughputs are comparable across topologies because we report absolute throughput (packets/node/ns).

We observe several expected trends in the results. 
The saturation throughput results are in precisely the same order 
as in our analytical expectation, with LPBT variants performing poorly and Kite being the best among the expert-designed networks across all sizes. 
\name outperforms expert-designed topologies at all scales and traffic types. 
For example, \name's large topologies optimized for latency and bandwidth achieve about 38\% higher saturation throughput than the best expert-designed topology for coherence traffic.
The Kite-Small topology which is identical to the \name small topology still achieves lower saturation throughput than \name because the simple heuristic NDBT routing results in lower saturation throughput than both variants of \name.
Finally, we note that for coherence traffic, the impact of the true cut-based bandwidth bounds on the Kite-Large, Double Butterfly, and Butter Donut topologies is visible in the reduced saturation throughput of these topologies.

\figputV{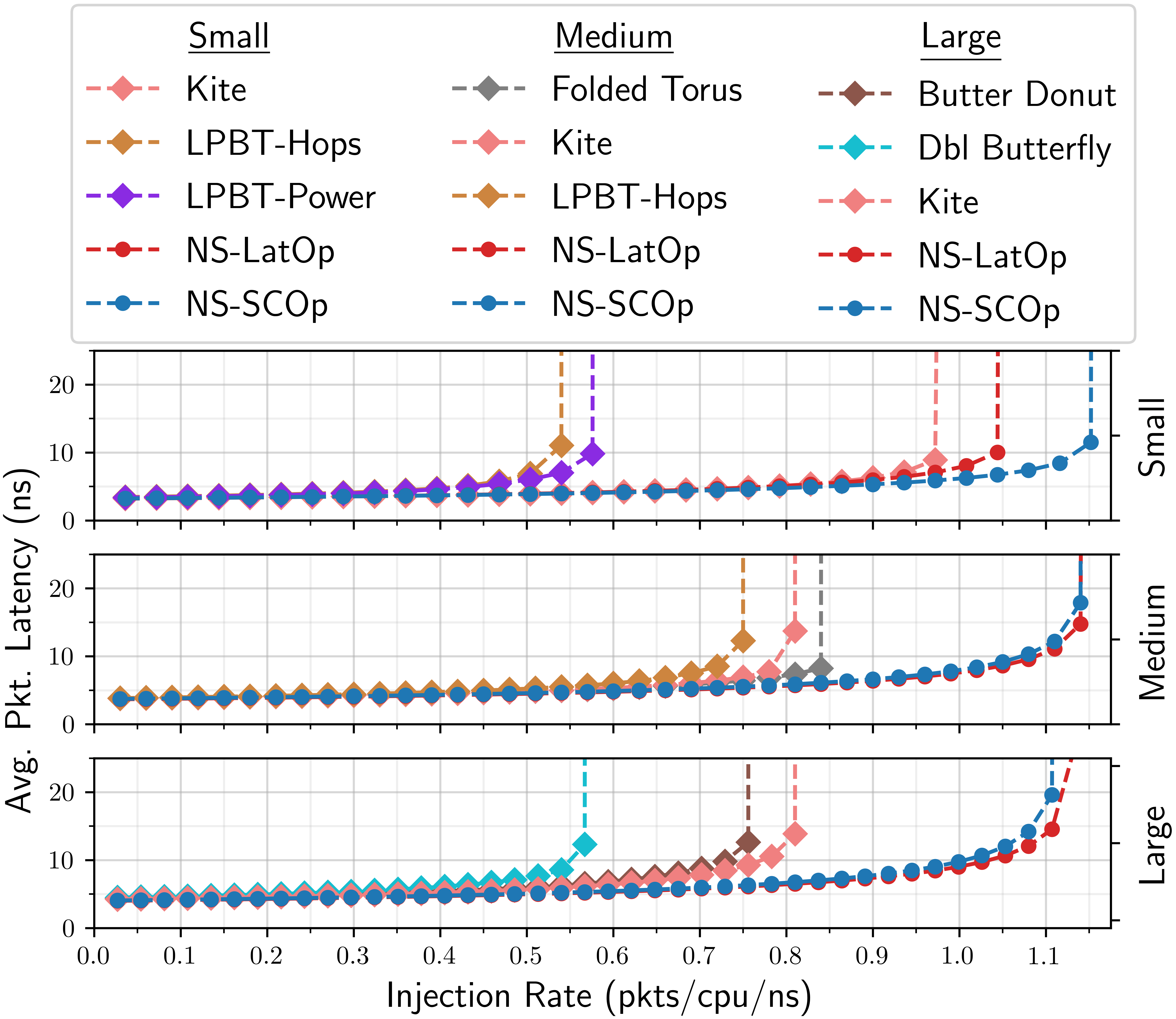}{Coherence traffic}{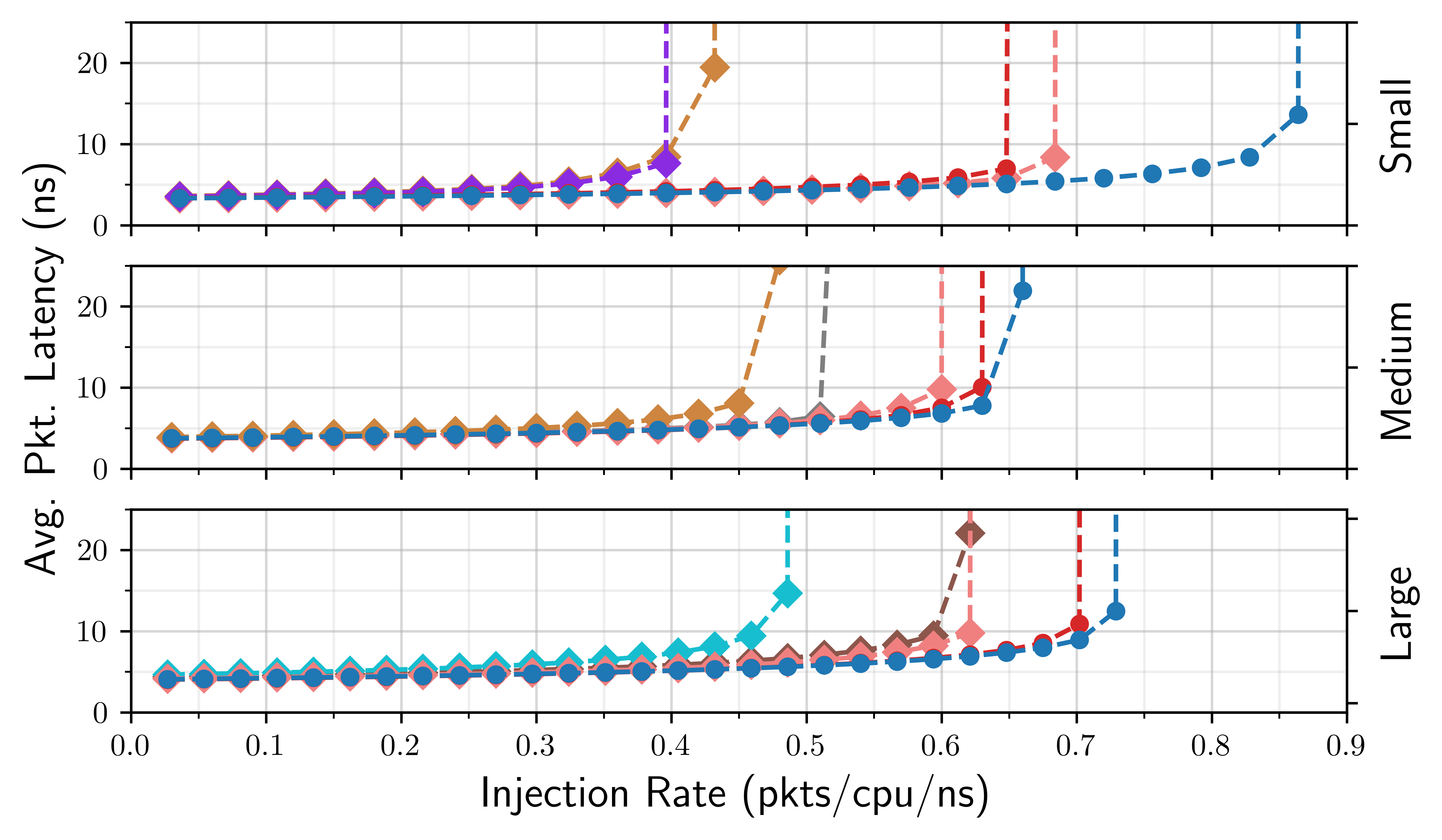}{Memory traffic}{Synthetic traffic with 20 (4x5) router NoIs}

\figref{main-flat_coh}(b) is a similar graph, but exercised with memory traffic instead.
The true-contention/
hot-spot behaviour is a tighter saturation bottleneck for 
memory traffic than the sparsest cut bandwidth. This is visible in the 
fact that {\em all } topologies saturate well beneath the traffic levels sustained with coherence traffic (note the difference in X-axis scales). 
Finally, the small topologies see a performance benefit with memory traffic because the higher clockspeed is able to achieve better throughput in terms of memory transactions, which in-turns dissipates the hotspot effect more effectively.

\putsubsec{isolation-results}{Isolating \name's topology and routing benefits}

\figput{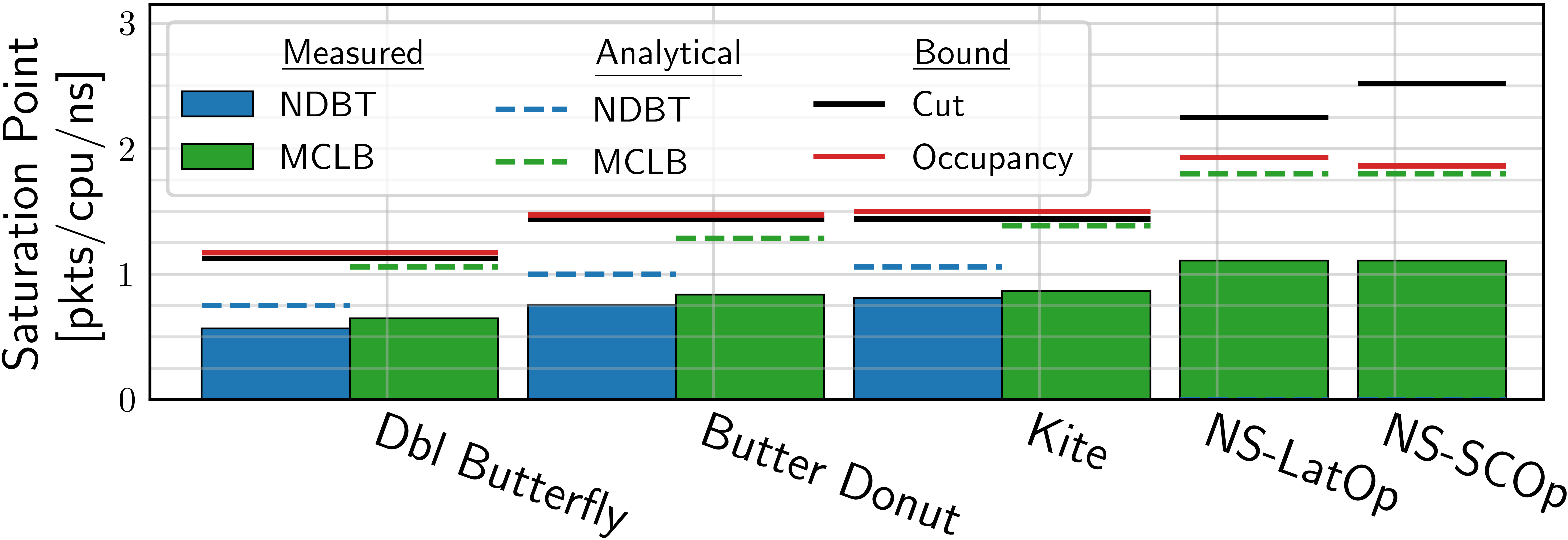}{0.9\linewidth}{Isolating \name's topology and routing benefits}

\name differs from legacy expert-designed topologies in two ways. In addition to the topology being generated by an optimization framework, 
the routing algorithm is also auto-generated to minimize maximum channel load (while retaining shortest path routing). 
To isolate the benefits of topology and routing, we evaluate legacy topologies with our MCLB routing. 

We limit this isolation study to large topologies due to lack of space; the results are qualitatively similar at other sizes.
\figref{isolations} shows the throughput achieved (Y-axis)  by various large topologies (X-axis) for NDBT and MCLB routing. (\name employs MCLB routing only). To compare the achieved throughput against expected analytical throughput, we also plot the ideal expected throughput (dashed lines). Finally, we also show the cut-based and occupancy-based saturation throughput bounds (solid lines) for each topology.

MCLB routing improves both observed and expected saturation throughput compared to heuristic routing, yet, still significantly lower than those equivalent values for \name-generated topologies.
MCLB-routing approaches the tighter of two upper bounds; cut-based for previous topologies, and occupancy-bound for those of \name.
Most apparent is the drastic difference in expected bounds. The occupancy bound is directly proportional to average hops as given in \tabref{metrics} and the cut bound is calculated from sparsest cut, both of which are known to be significantly greater for \name-generated topologies.
Note that the gap between analytical expectation  and measured throughput is due to well-known challenges of input-queued networks as analyzed in \cite{karol1987inputversusoutputqueueing}.

\putsubsec{fullsystem-results}{Real Workload Simulation}

\figputW{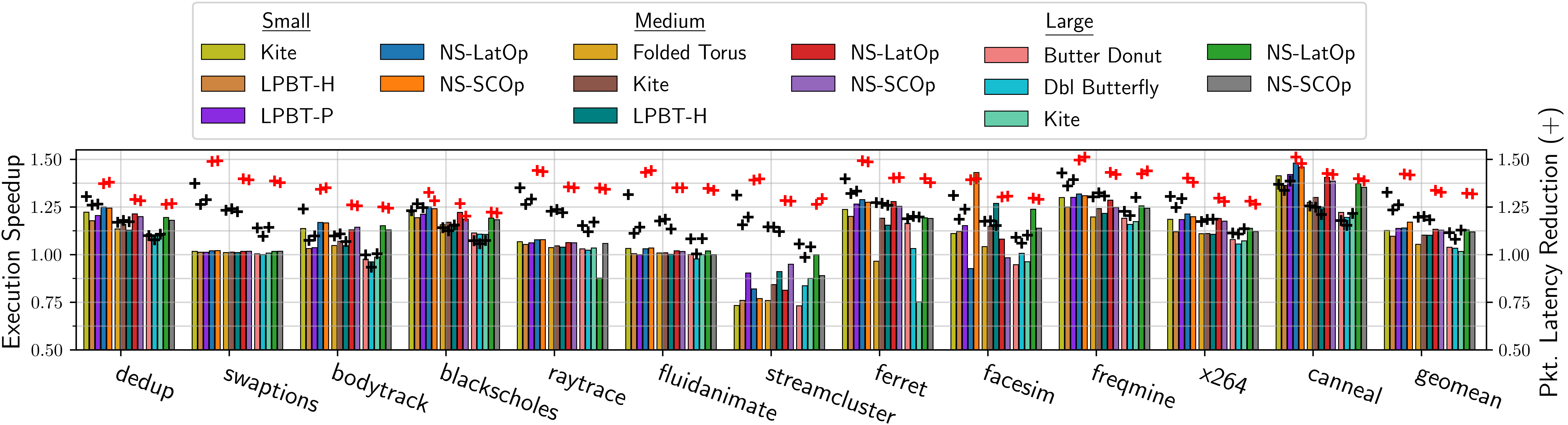}{0.8\linewidth}{PARSEC application speedup and packet latency reduction}

Network topologies directly affect packet latency, and indirectly affect application execution time by speeding up coherence and memory traffic. Our evaluation aims to show both these effects of \name-generated topologies.
The simulated PARSEC benchmarks and the calculated geometric mean are listed on the X-axis of \figref{parsec-flat} given in increasing order of L2 misses per instruction -- related to impact of network performance on computation -- from left to right.
The execution time speedup (left Y-axis) in simulated cycles is presented as a bar chart, grouped according to the small, medium and large classifications. The reduction in average packet latency (right Y-axis) relative to
mesh is shown using markers with \name topologies are highlighted in red for easy identification.

Across the board we see that packet latencies see improvements
with \name topologies always yielding the highest reduction in latency. Comparing packet latency reductions to application performance, we see that there is broad correlation (higher packet latency reductions correlate with higher performance) while
the sensitivity of application performance to network packet latency performance differs from benchmark to benchmark.
 
The topology with the greatest speedup and packet latency reduction was NS-LatOp-small 19\% speedup in execution time and 33\% reduction in packet latency relative to mesh. As a ``small" topology, NS-LatOp gains outperforms comparable ``small" topologies (which all use the same high clockspeeds).
Across the board, the LPBT topologies see the worse speedups and packet latency reductions.
With the medium topologies, Folded Torus has the worst performance. NS-LatOp-Medium and NS-SCOp-Medium both outperform Kite-Medium by 20\% speedup. The greatest improvements of \name over expert-designed is for large where NS-LatOp-large has a 11\% speedup over Kite Large.

\putsubsec{dsent}{Area and Power Analysis}

\figput{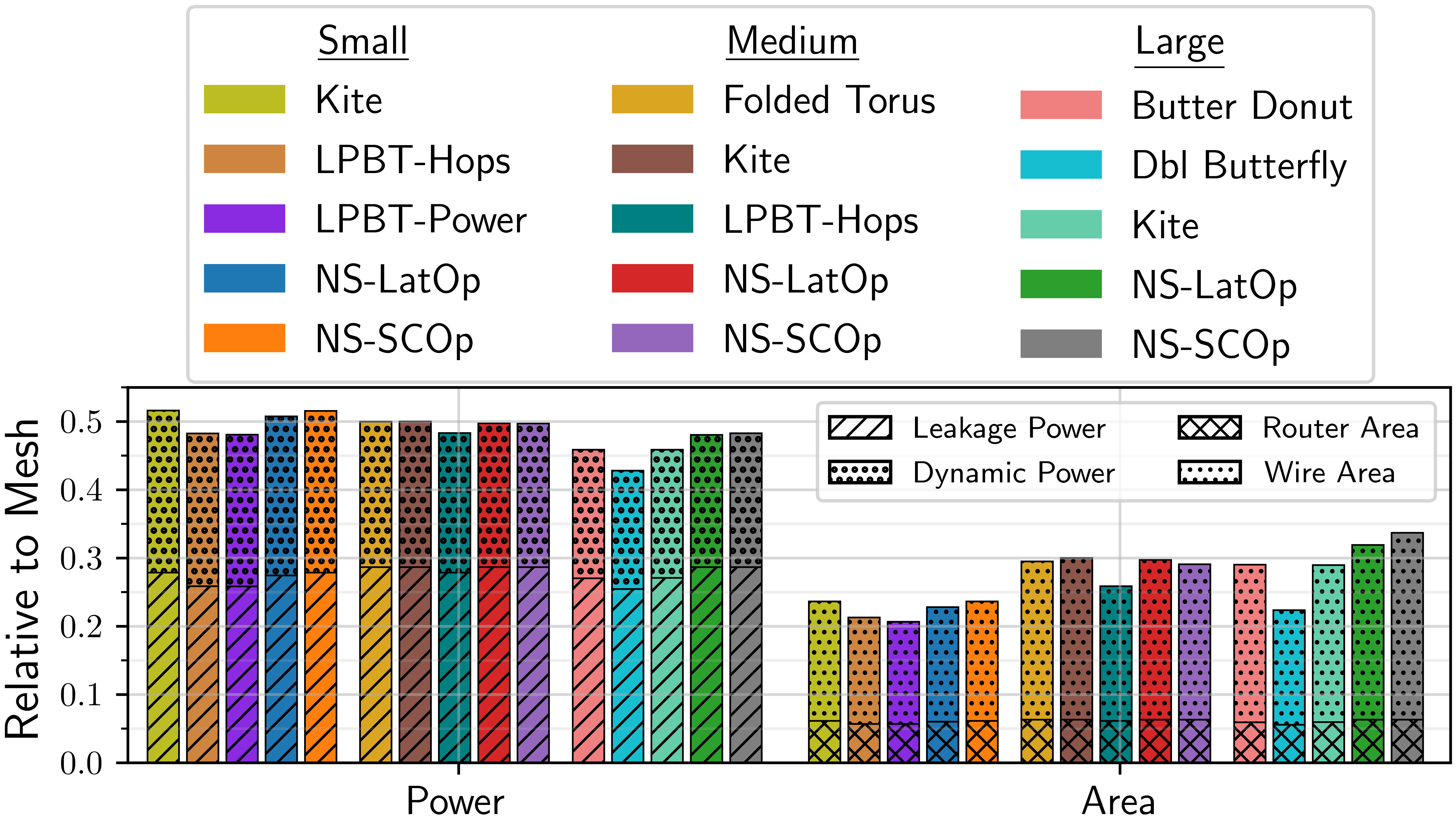}{}{Power and area relative to mesh}

We examine the area and power impacts using the 22nm bulk LVT technology model of DSENT \cite{sun2012dsent}.
Activity statistics on just the NoI topology was input to DSENT to calculate the dynamic and leakage power consumed. The values were normalized to a mesh topology (lower is better) to provide a common baseline and are given in \figref{dsent}. The power consumption of each topology is shown as a stacked bar with static and dynamic sub-bars.
The leakage power is more or less the same across all the topologies because they all have 20 routers and similar number of links. Further, the leakage is comparable to the dynamic power consumption, as expected \cite{guang2009reviewofpowernoc}. The dynamic power has a variable component because the aggregate wire-lengths vary across topologies.
\name large topologies have approximately 17\% lower dynamic power usage than their equivalent small topologies, resulting in 7\% lower total power; recall that this is expected because large topologies use slower clocks than the small topologies.
\figref{dsent} also shows the normalized area of the NoI. The area of routers and wires are shown separately to highlight the relative size of wires to routers. The total wire area is the dominant fraction as all topologies use the same routers with the same radix. Naturally, \name's maximizes the use of links to achieve better performance, and this incurs a higher cost. (Note that \name topologies are under 3\% of interposer area; the interposer remains sparse and ``minimally-active''.)

\putsubsec{scalability-results}{Traffic Pattern and Scalablity Synthetic Evaluations}

\figput{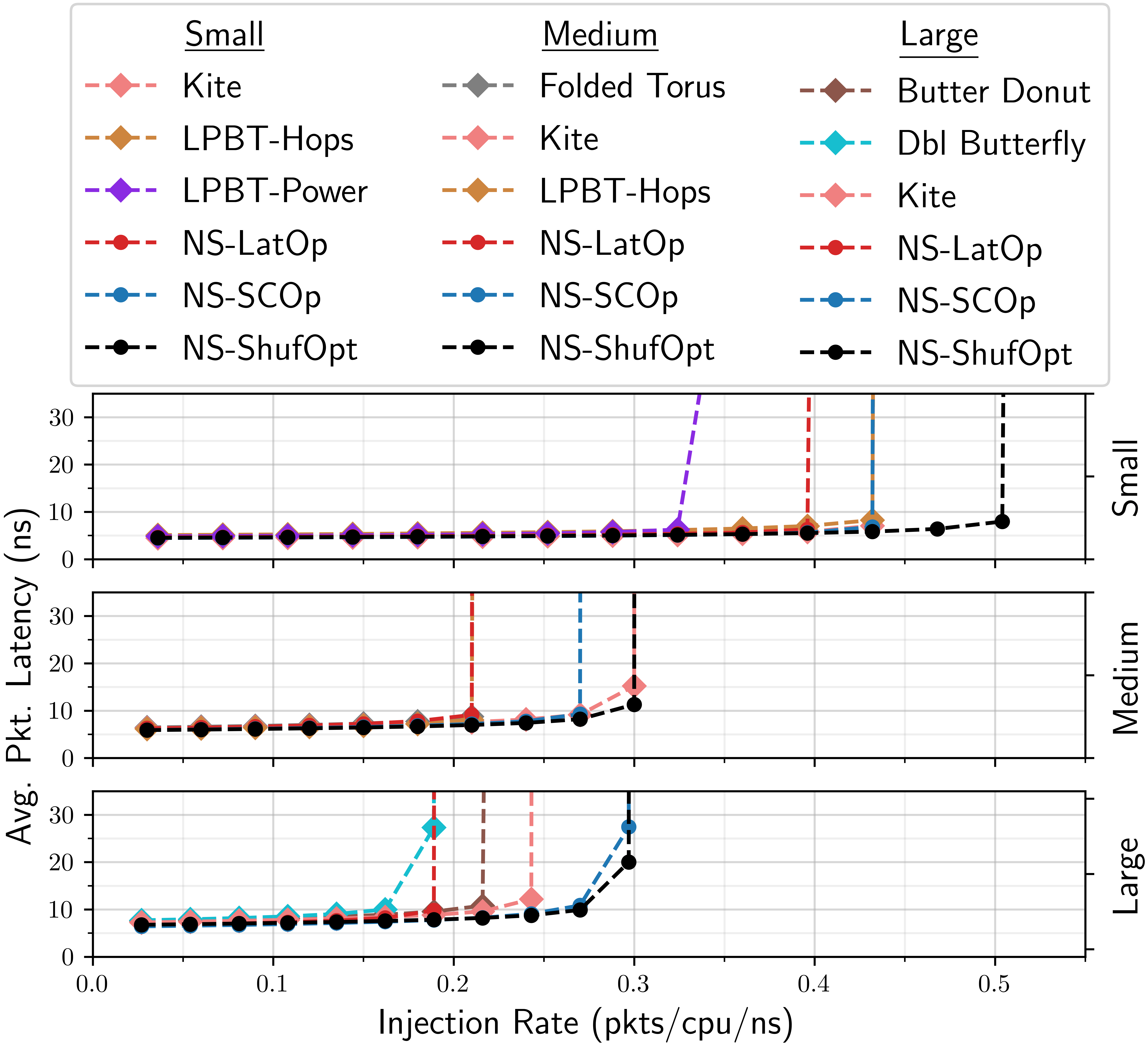}{0.9\linewidth}{Synthetic shuffle traffic pattern on shuffle-optimized topologies}

To showcase \name's versatility outside uniform random, we used it to generate topologies optimized for basic gem5 traffic pattern, ``shuffle," because it has far source-destination pairs and applies to the 5x4 layout \cite{gem5}. These pattern-optimized topologies are labeled as ``NS\_ShufOpt" in subsequent graphs.
iven the numbering in \figref{ns_m_latop} and the number of routers $n$, the source and destination pairs are calculated as follows. 

% pattern is...
% if (source < num_destinations/2)
%     destination = source*2;
% else
%     destination = (source*2 - num_destinations + 1);

\vspace{0.3em}
$dest = \begin{cases}
    2src  & 0 \leq src \textless \frac{n}{2}\\
    (2src + 1) mod\ n & \frac{n}{2} \leq src \textless n
\end{cases}$
\vspace{0.3em}

\figref{shuffle-flat_coh} shows the latency (Y-axis) and throughput (X-axis) achieved under the shuffle traffic pattern for the small, medium and large networks with the added ``NS\_ShufOpt" in black.
The legacy networks and \name networks optimized for random traffic exhibit varying behavior as they are not customized for the specific shuffle pattern.
However, the shuffle-optimized \name topology outperforms all other solutions.

\figput{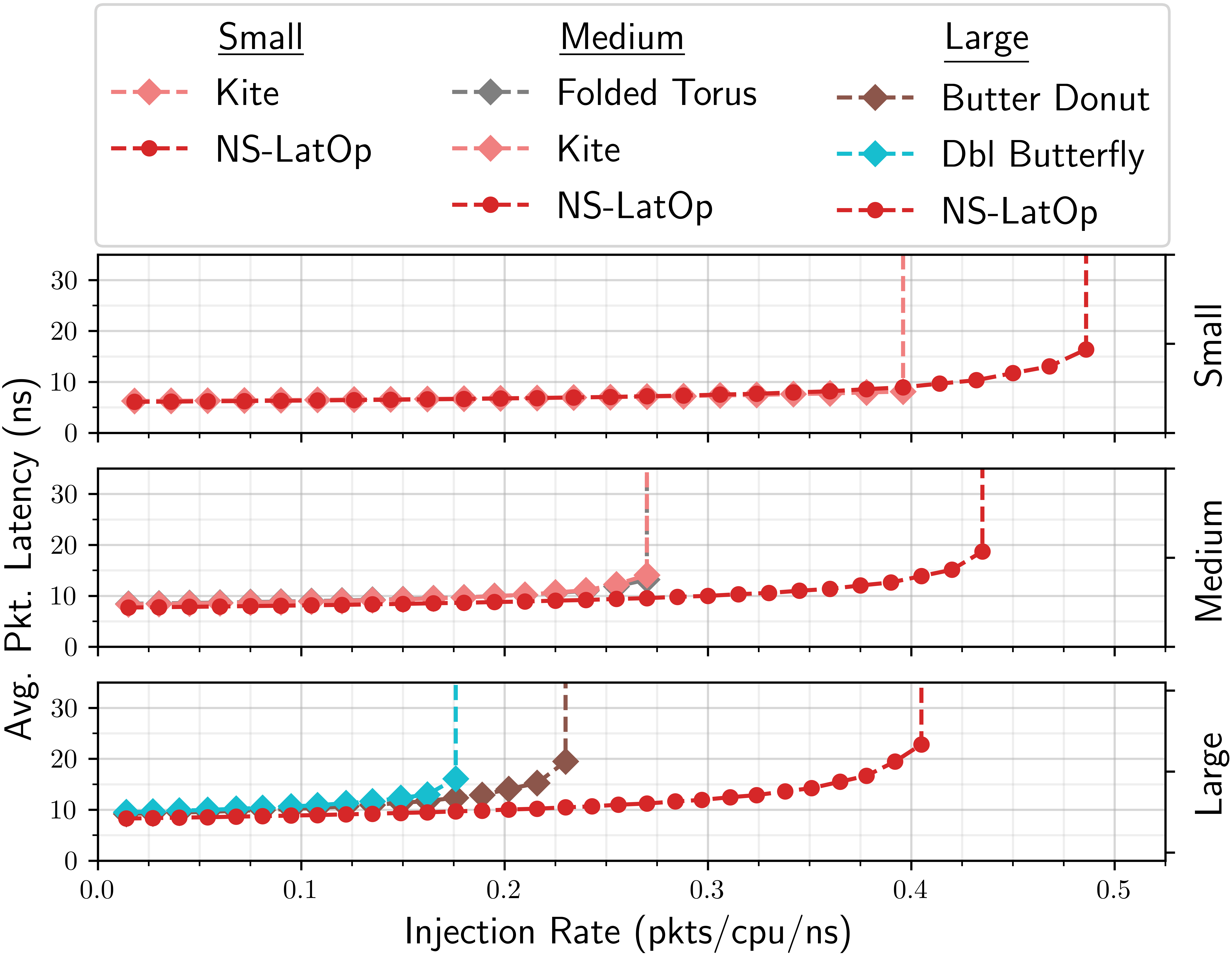}{0.9\linewidth}{Synthetic traffic with 48 (8x6) router NoIs}

To empirically evaluate \name's ability to scale, we simulated a larger 48-router network (with 8x6 placement) with synthetic traffic and found that \name's topologies continue to outperform expert-designed topologies. 
We compare against a subset of topologies that we were able to scale up to 48 nodes according to the rules of the topology \cite{butterdonut,kite}. Kite-Large could not be scaled as it seems to rely on an odd number of columns for symmetry. The LPBT was unable to generate a connected/sufficient graph.
Due to lack of space, we only show results focusing on saturation throughput with synthetic uniform random traffic.

\figref{48r-flat_coh} shows the latency (Y-axis) and throughput (X-axis) achieved for the small, medium and large networks. Across all the link-length categories, \name-generated topologies far outperform the legacy topologies in saturation throughput with 18\%, 56\%, and 67\% higher throughput than the best competing small, medium, and large topologies, respectively. Note that \name uses latency-optimized topologies and yet achieves significantly better throughput as well.

\putsec{related}{Related Work}

Given the extensive discussion of related interposer network designs in the body of the paper, we focus the discussion in this section to related work from the NoC/SoC space. Such related work may be broadly categorized into the following categories.

\paragraph{MILP Optimization to exploit known application-specific communication}

For application specific communication needs, there are existing MIP-based approaches to minimize link contention (also known as channel load) via custom routing on fixed topologies~\cite{abdelgawad2011transcom}.
Unlike \name, they do not generate new topologies.
Alternately, there are similar approaches that use the known communication graph between IP cores to synthesize custom NoCs via placement and SoC floor-planning~\cite{lpbt,chatha2008automated,li2011floorplanning,huang2012applicationspecific,zhong2011applicationspecific,mukherjee2016ilpfloorplan}. The synthesis objectives are variously to minimize power, minimize resources (area) or to fit with power/area requirements. We have compared against \cite{lpbt} (LPBT) in both analytical results and experimental results where \name is shown to be faster to solve {\em and} produces higher performance networks.
The optimization formulation of ~\cite{lpbt} defined hops along a path through a series of port mapping variables from each router. Therefore, the formulation necessarily computes the path of every single source and destination during the MIP solver. This is unlike \name, which defines path lengths as distances to other known distances (C4 in \tabref{mip_form})
and leaves the discovery of shortest paths to post-solver evaluations. This latency formulation is novel amongst all related works~\cite{lpbt,chatha2008automated,huang2012applicationspecific,zhong2011applicationspecific,mukherjee2016ilpfloorplan}.
Some work attempts to produce specific topologies (e.g., Steiner trees~\cite{gangwar2020automatedsynthesis}) in the SoC context which have poor cut-based bandwidth bounds  and thus are unsuitable for general purpose multicores. 
None of them solve the problem of maximizing throughput nor even proxy metrics like bisection bandwidth. Note that the use of solvers is merely a commonality of the tool; the difference in formulation, constraints and objectives reveals that \name solves a different problem than the aforementioned works.

\paragraph{Algorithmic optimization}
Many works design algorithms to improve fault-tolerance ~\cite{tosun2012applicationspecific,tosun2015faulttolerant,bezerra2019tabu} 
or power/area ~\cite{chang2008lowpower,joseph2019area,choudhary2010genetic}. However, none of these works define and optimize for latency (hop count) or bandwidth.

\paragraph{Resource placement}
Vaish \etal ~\cite{vaish2016optimization} use MIP to optimize resource placement problems (e.g., placement of optional memory controllers, additional virtual channels, and wider physical links) in a regular mesh topology. Le \etal~\cite{le2017optimizing} use MIP optimization to place a minimal set of hybrid router resources for routing among ``subnets" in a manycore NoC, based on application-specific knowledge. \name's goal is to define the topology, not to selectively place/map resources in it. 
Dumitriu and Khan~\cite{dumitriu2009throughput} seek to reduce resource usage while maintaining certain latency-based quality of service guarantees.
Other resource allocation solutions use carefully-designed algorithms ~\cite{bertozzi2005synthesis,murali2005mapping,murali2006designingnoc,le2017reconfigurable,phanibhushana2014nocdesign}.

\putsec{conclu}{Conclusion}

Recent interest in NoI topology design has produced a series of ever improving networks~\cite{kite,butterdonut,lpbt}. Given that the high-level goals that expert designers strive for -- low average latency, high saturation bandwidth, and cost-efficiency -- are well-known, the natural question is whether network design can be automated to maximally achieve these goals. Prior approaches to automated network design from the embedded domain (i.e., NoC synthesis for SoCs)
are inadequate solutions for general purpose multicores; we show that such synthesized networks
underperform expert-designed topologies.
Our contribution -- a novel optimization-based, topology generation framework called \name -- demonstrates that it is possible to design high-performance NoI topologies for general-purpose multicores that achieve better latency and throughput than legacy networks while satisfying cost constraints and with reasonable solver times.
The generated topologies are compatible with automatic, high-performance deadlock-free routing techniques and VC allocation techniques for irregular networks. 

%Further, routing and VC allocation strategies may also be automatically generated to maximize throughput.   
%The superior analytical metrics of \name-generated topologies can be effectively implemented through automatic, established routing and VC allocation schemes.
% Improving the latency and throughput metrics does not incur implementation cost as the irregular topologies can be effectively implemented through automatic, established routing and VC allocation schemes.

\name's generated topologies reveal several desirable features.
Some of them are optimal in the objective function; no expert can do better. And even the topologies not known to be optimal handily outperform expert-designed topologies in analytical metrics (e.g., 13.5\% lower average hops and 75\% higher saturation bandwidth).
Simulations show that
\name-generated topologies achieve  mean speedups of up to 11\% for PARSEC workloads over prior topologies.
Finally, \name's computation effort is not overwhelming; the generated topologies shown in this paper were all generated in minutes/hours of solver time.

\bibliographystyle{plain}
% \bibliography{local}
\bibliography{paper}

\begin{thebibliography}{10}

\bibitem{abdelgawad2011transcom}
Ahmed~H. Abdel-Gawad and Mithuna Thottethodi.
\newblock Transcom: Transforming stream communication for load balance and efficiency in networks-on-chip.
\newblock In {\em Proceedings of the 44th Annual IEEE/ACM International Symposium on Microarchitecture}, MICRO-44, page 237–247, New York, NY, USA, 2011. Association for Computing Machinery.

\bibitem{bartolini2005recent}
Sandro Bartolini, Roberto Giorgi, Enrico Martinelli, and Zdravko Popovic.
\newblock Recent proposals for tiled architectures.
\newblock {\em Poster Abstract of the HiPEAC ACACES-2005 Summer School}, pages 47--50, 2005.

\bibitem{bertozzi2005synthesis}
D.~Bertozzi, A.~Jalabert, Srinivasan Murali, R.~Tamhankar, S.~Stergiou, L.~Benini, and G.~De~Micheli.
\newblock Noc synthesis flow for customized domain specific multiprocessor systems-on-chip.
\newblock {\em IEEE Transactions on Parallel and Distributed Systems}, 16(2):113--129, 2005.

\bibitem{bezerra2019tabu}
Gustavo~Alves Bezerra, Patrícia~Pontes Cruz, Márcio~Eduardo Kreutz, and Monica~Magalhães Pereira.
\newblock Generation of application specific fault tolerant irregular noc topologies using tabu search.
\newblock In {\em 2019 IX Brazilian Symposium on Computing Systems Engineering (SBESC)}, pages 1--8, 2019.

\bibitem{kite}
Srikant Bharadwaj, Jieming Yin, Bradford Beckmann, and Tushar Krishna.
\newblock Kite: A family of heterogeneous interposer topologies enabled via accurate interconnect modeling.
\newblock In {\em 2020 57th ACM/IEEE Design Automation Conference (DAC)}, pages 1--6, 2020.

\bibitem{bienia2008parsec}
Christian Bienia, Sanjeev Kumar, Jaswinder~Pal Singh, and Kai Li.
\newblock The parsec benchmark suite: Characterization and architectural implications.
\newblock In {\em 2008 International Conference on Parallel Architectures and Compilation Techniques (PACT)}, pages 72--81, 2008.

\bibitem{isidor2023disintegratingmanycores}
Isidor~R. Brki\'{c} and Mark~C. Jeffrey.
\newblock Disintegrating manycores: Which applications lose and why?
\newblock In {\em Proceedings of the 16th International Workshop on Network on Chip Architectures}, NoCArc '23, page 3–8, New York, NY, USA, 2023. Association for Computing Machinery.

\bibitem{capone2010routingschedulingandchannelassignment}
A.~Capone, G.~Carello, I.~Filippini, S.~Gualandi, and F.~Malucelli.
\newblock Routing, scheduling and channel assignment in wireless mesh networks: Optimization models and algorithms.
\newblock {\em Ad Hoc Networks}, 8(6):545--563, 2010.

\bibitem{chang2008lowpower}
K.-C. Chang.
\newblock Low-power algorithm for automatic topology generation for application-specific networks on chips.
\newblock {\em IET Computers \& Digital Techniques}, 2:239--249(10), May 2008.

\bibitem{chatha2008automated}
Karam~S. Chatha, Krishnan Srinivasan, and Goran Konjevod.
\newblock Automated techniques for synthesis of application-specific network-on-chip architectures.
\newblock {\em IEEE Transactions on Computer-Aided Design of Integrated Circuits and Systems}, 27(8):1425--1438, 2008.

\bibitem{choudhary2010genetic}
Naveen Choudhary, M.~S. Gaur, V.~Laxmi, and V.~Singh.
\newblock Genetic algorithm based topology generation for application specific network-on-chip.
\newblock In {\em Proceedings of 2010 IEEE International Symposium on Circuits and Systems}, pages 3156--3159, 2010.

\bibitem{vanchu2019lef}
Thiem~Van CHU and Kenji KISE.
\newblock Lef: An effective routing algorithm for two-dimensional meshes.
\newblock {\em IEICE Transactions on Information and Systems}, E102.D(10):1925--1941, 2019.

\bibitem{clr}
Thomas~H. Cormen, Charles~E. Leiserson, Ronald~L. Rivest, and Clifford Stein.
\newblock Introduction to algorithms, second edition.
\newblock 2001.

\bibitem{dally1988deadlock}
William~J Dally and Charles~L Seitz.
\newblock Deadlock-free message routing in multiprocessor interconnection networks.
\newblock 1988.

\bibitem{domke2011deadlock}
Jens Domke, Torsten Hoefler, and Wolfgang~E. Nagel.
\newblock Deadlock-free oblivious routing for arbitrary topologies.
\newblock In {\em 2011 IEEE International Parallel \& Distributed Processing Symposium}, pages 616--627, 2011.

\bibitem{duato}
J.~Duato.
\newblock A necessary and sufficient condition for deadlock-free adaptive routing in wormhole networks.
\newblock {\em IEEE Transactions on Parallel and Distributed Systems}, 6(10):1055--1067, 1995.

\bibitem{dumitriu2009throughput}
Victor Dumitriu and Gul~N. Khan.
\newblock Throughput-oriented noc topology generation and analysis for high performance socs.
\newblock {\em IEEE Transactions on Very Large Scale Integration (VLSI) Systems}, 17(10):1433--1446, 2009.

\bibitem{ebrahimi2017ebda}
Masoumeh Ebrahimi and Masoud Daneshtalab.
\newblock Ebda: A new theory on design and verification of deadlock-free interconnection networks.
\newblock ISCA '17, page 703–715, New York, NY, USA, 2017. Association for Computing Machinery.

\bibitem{gangwar2020automatedsynthesis}
Anup Gangwar, Nitin~Kumar Agarwal, Ravishankar Sreedharan, Ambica Prasad, Sri~Harsha Gade, and Zheng Xu.
\newblock Automated synthesis of custom networks-on-chip for real world applications.
\newblock In {\em Proceedings of the 39th International Conference on Computer-Aided Design}, ICCAD '20, New York, NY, USA, 2020. Association for Computing Machinery.

\bibitem{guang2009reviewofpowernoc}
Liang Guang, Pasi Liljeberg, Ethiopia Nigussie, and Hannu Tenhunen.
\newblock A review of dynamic power management methods in noc under emerging design considerations.
\newblock In {\em 2009 NORCHIP}, pages 1--6, 2009.

\bibitem{gurobi}
{Gurobi Optimization, LLC}.
\newblock {Gurobi Optimizer Reference Manual}, 2023.

\bibitem{hartmanis1982miphard}
Juris Hartmanis.
\newblock Computers and intractability: A guide to the theory of np-completeness (michael r. garey and david s. johnson).
\newblock {\em SIAM Review}, 24(1):90--91, 1982.

\bibitem{hu2018diestacking}
Xing Hu, Dylan Stow, and Yuan Xie.
\newblock Die stacking is happening.
\newblock {\em IEEE Micro}, 38(1):22--28, 2018.

\bibitem{huang2012applicationspecific}
Bo~Huang, Song Chen, Wei Zhong, and Takeshi Yoshimura.
\newblock Application-specific network-on-chip synthesis with topology-aware floorplanning.
\newblock In {\em 2012 25th Symposium on Integrated Circuits and Systems Design (SBCCI)}, pages 1--6, 2012.

\bibitem{doublebutterfly}
Natalie D.~Enright Jerger, Ajaykumar Kannan, Zimo Li, and Gabriel~H. Loh.
\newblock Noc architectures for silicon interposer systems: Why pay for more wires when you can get them (from your interposer) for free?
\newblock {\em 2014 47th Annual IEEE/ACM International Symposium on Microarchitecture}, pages 458--470, 2014.

\bibitem{joseph2019area}
Jan~Moritz Joseph, Dominik Ermel, Tobias Drewes, Lennart Bamberg, Alberto García-Oritz, and Thilo Pionteck.
\newblock Area optimization with non-linear models in core mapping for system-on-chips.
\newblock In {\em 2019 8th International Conference on Modern Circuits and Systems Technologies (MOCAST)}, pages 1--4, 2019.

\bibitem{jyoth2016measuringthroughput}
Sangeetha~Abdu Jyothi, Ankit Singla, P.~Brighten Godfrey, and Alexandra Kolla.
\newblock Measuring and understanding throughput of network topologies.
\newblock In {\em SC '16: Proceedings of the International Conference for High Performance Computing, Networking, Storage and Analysis}, pages 761--772, 2016.

\bibitem{butterdonut}
Ajaykumar Kannan, Natalie D.~Enright Jerger, and Gabriel~H. Loh.
\newblock Enabling interposer-based disintegration of multi-core processors.
\newblock {\em 2015 48th Annual IEEE/ACM International Symposium on Microarchitecture (MICRO)}, pages 546--558, 2015.

\bibitem{karol1987inputversusoutputqueueing}
M.~Karol, M.~Hluchyj, and S.~Morgan.
\newblock Input versus output queueing on a space-division packet switch.
\newblock {\em IEEE Transactions on Communications}, 35(12):1347--1356, 1987.

\bibitem{kinsy2009deadlock}
Michel~A. Kinsy, Myong~Hyon Cho, Tina Wen, Edward Suh, Marten van Dijk, and Srinivas Devadas.
\newblock Application-aware deadlock-free oblivious routing.
\newblock In {\em Proceedings of the 36th Annual International Symposium on Computer Architecture}, ISCA '09, page 208–219, New York, NY, USA, 2009. Association for Computing Machinery.

\bibitem{lau2021chiplet}
John~H. Lau.
\newblock {\em Chiplet Heterogeneous Integration}, pages 413--439.
\newblock Springer Singapore, Singapore, 2021.

\bibitem{le2017optimizing}
Tung Le, Rui Ning, Dan Zhao, Hongyi Wu, and Magdy Bayoumi.
\newblock Optimizing the heterogeneous network on-chip design in manycore architectures.
\newblock pages 184--189, 09 2017.

\bibitem{le2017reconfigurable}
Tung~Thanh Le, Dan Zhao, and Magdy Bayoumi.
\newblock Efficient reconfigurable global network-on-chip designs towards heterogeneous cpu-gpu systems: An application-aware approach.
\newblock In {\em 2017 IEEE Computer Society Annual Symposium on VLSI (ISVLSI)}, pages 439--444, 2017.

\bibitem{li2011floorplanning}
Katherine Shu-Min Li, Shu-Yu Chen, Liang-Bi Chen, and Ruei-Ting Gu.
\newblock A fast custom network topology generation with floorplanning for noc-based systems.
\newblock In {\em 2011 IEEE International Conference on IC Design \& Technology}, pages 1--4, 2011.

\bibitem{gem5}
Jason Lowe{-}Power, Abdul~Mutaal Ahmad, Ayaz Akram, Mohammad Alian, Rico Amslinger, Matteo Andreozzi, Adri{\`{a}} Armejach, Nils Asmussen, Srikant Bharadwaj, Gabe Black, Gedare Bloom, Bobby~R. Bruce, Daniel~Rodrigues Carvalho, Jer{\'{o}}nimo Castrill{\'{o}}n, Lizhong Chen, Nicolas Derumigny, Stephan Diestelhorst, Wendy Elsasser, Marjan Fariborz, Amin~Farmahini Farahani, Pouya Fotouhi, Ryan Gambord, Jayneel Gandhi, Dibakar Gope, Thomas Grass, Bagus Hanindhito, Andreas Hansson, Swapnil Haria, Austin Harris, Timothy Hayes, Adrian Herrera, Matthew Horsnell, Syed Ali~Raza Jafri, Radhika Jagtap, Hanhwi Jang, Reiley Jeyapaul, Timothy~M. Jones, Matthias Jung, Subash Kannoth, Hamidreza Khaleghzadeh, Yuetsu Kodama, Tushar Krishna, Tommaso Marinelli, Christian Menard, Andrea Mondelli, Tiago M{\"{u}}ck, Omar Naji, Krishnendra Nathella, Hoa Nguyen, Nikos Nikoleris, Lena~E. Olson, Marc~S. Orr, Binh Pham, Pablo Prieto, Trivikram Reddy, Alec Roelke, Mahyar Samani, Andreas Sandberg, Javier Setoain, Boris Shingarov, Matthew~D.
  Sinclair, Tuan Ta, Rahul Thakur, Giacomo Travaglini, Michael Upton, Nilay Vaish, Ilias Vougioukas, Zhengrong Wang, Norbert Wehn, Christian Weis, David~A. Wood, Hongil Yoon, and {\'{E}}der~F. Zulian.
\newblock The gem5 simulator: Version 20.0+.
\newblock {\em CoRR}, abs/2007.03152, 2020.

\bibitem{lynse2006layeredrouting}
O.~Lysne, T.~Skeie, S.-A. Reinemo, and I.~Theiss.
\newblock Layered routing in irregular networks.
\newblock {\em IEEE Transactions on Parallel and Distributed Systems}, 17(1):51--65, 2006.

\bibitem{mukherjee2016ilpfloorplan}
Priyajit Mukherjee and Santanu Chattopadhyay.
\newblock An ilp-based floorplan-aware path synthesis technique for application-specific noc design.
\newblock In {\em 2016 3rd International Conference on Recent Advances in Information Technology (RAIT)}, pages 543--548, 2016.

\bibitem{murali2005mapping}
Srinivasan Murali, Luca Benini, and Giovanni De~Micheli.
\newblock Mapping and physical planning of networks-on-chip architectures with quality-of-service guarantees.
\newblock In {\em Proceedings of the 2005 Asia and South Pacific Design Automation Conference}, ASP-DAC '05, page 27–32, New York, NY, USA, 2005. Association for Computing Machinery.

\bibitem{murali2006designingnoc}
Srinivasan Murali, Paolo Meloni, Federico Angiolini, David Atienza, Salvatore Carta, Luca Benini, Giovanni De~Micheli, and Luigi Raffo.
\newblock Designing application-specific networks on chips with floorplan information.
\newblock In {\em 2006 IEEE/ACM International Conference on Computer Aided Design}, pages 355--362, 2006.

\bibitem{naffziger2021chiplet}
Samuel Naffziger, Noah Beck, Thomas Burd, Kevin Lepak, Gabriel~H. Loh, Mahesh Subramony, and Sean White.
\newblock Pioneering chiplet technology and design for the amd epyc™ and ryzen™ processor families : Industrial product.
\newblock In {\em 2021 ACM/IEEE 48th Annual International Symposium on Computer Architecture (ISCA)}, pages 57--70, 2021.

\bibitem{pascual2018deadlockavoidanceforarbitrary}
Jose~A. Pascual and Javier Navaridas.
\newblock High-performance, low-complexity deadlock avoidance for arbitrary topologies/routings.
\newblock In {\em Proceedings of the 2018 International Conference on Supercomputing}, ICS '18, page 129–138, New York, NY, USA, 2018. Association for Computing Machinery.

\bibitem{phanibhushana2014nocdesign}
Bharath Phanibhushana and Sandip Kundu.
\newblock Network-on-chip design for heterogeneous multiprocessor system-on-chip.
\newblock In {\em 2014 IEEE Computer Society Annual Symposium on VLSI}, pages 486--491, 2014.

\bibitem{schwiebert2001deadlock}
L.~Schwiebert.
\newblock Deadlock-free oblivious wormhole routing with cyclic dependencies.
\newblock {\em IEEE Transactions on Computers}, 50(9):865--876, 2001.

\bibitem{shim2023NCDE}
Jae~Eun Shim, Mingu Kang, and Tae~Hee Han.
\newblock Ncde: In-network caching for directory entries to expedite data access in tiled-chip multiprocessors.
\newblock {\em IEEE Access}, 11:3080--3095, 2023.

\bibitem{sodani2016knightslanding}
Avinash Sodani, Roger Gramunt, Jesus Corbal, Ho-Seop Kim, Krishna Vinod, Sundaram Chinthamani, Steven Hutsell, Rajat Agarwal, and Yen-Chen Liu.
\newblock Knights landing: Second-generation intel xeon phi product.
\newblock {\em IEEE Micro}, 36(2):34--46, 2016.

\bibitem{lpbt}
K.~Srinivasan, K.S. Chatha, and G.~Konjevod.
\newblock Linear-programming-based techniques for synthesis of network-on-chip architectures.
\newblock {\em IEEE Transactions on Very Large Scale Integration (VLSI) Systems}, 14(4):407--420, 2006.

\bibitem{stewart2006kautz}
Lawrence~C Stewart and David Gingold.
\newblock A new generation of cluster interconnect.
\newblock {\em White Paper, SiCortex Inc}, 2006.

\bibitem{stow2017costeffective}
Dylan Stow, Yuan Xie, Taniya Siddiqua, and Gabriel~H. Loh.
\newblock Cost-effective design of scalable high-performance systems using active and passive interposers.
\newblock In {\em 2017 IEEE/ACM International Conference on Computer-Aided Design (ICCAD)}, pages 728--735, 2017.

\bibitem{sun2012dsent}
Chen Sun, Chia-Hsin~Owen Chen, George Kurian, Lan Wei, Jason Miller, Anant Agarwal, Li-Shiuan Peh, and Vladimir Stojanovic.
\newblock Dsent - a tool connecting emerging photonics with electronics for opto-electronic networks-on-chip modeling.
\newblock In {\em 2012 IEEE/ACM Sixth International Symposium on Networks-on-Chip}, pages 201--210, 2012.

\bibitem{tosun2012applicationspecific}
S.~Tosun.
\newblock Application-specific topology generation algorithms for network-on-chip design.
\newblock {\em IET Computers \& Digital Techniques}, 6:318--333(15), September 2012.

\bibitem{tosun2015faulttolerant}
Suleyman Tosun, Vahid~B. Ajabshir, Ozge Mercanoglu, and Ozcan Ozturk.
\newblock Fault-tolerant topology generation method for application-specific network-on-chips.
\newblock {\em IEEE Transactions on Computer-Aided Design of Integrated Circuits and Systems}, 34(9):1495--1508, 2015.

\bibitem{vaish2016optimization}
Nilay Vaish, Michael~C. Ferris, and David~A. Wood.
\newblock Optimization models for three on-chip network problems.
\newblock {\em ACM Trans. Archit. Code Optim.}, 13(3), sep 2016.

\bibitem{yin2018modularrouting}
Jieming Yin, Zhifeng Lin, Onur Kayiran, Matthew Poremba, Muhammad Shoaib Bin~Altaf, Natalie Enright~Jerger, and Gabriel~H. Loh.
\newblock Modular routing design for chiplet-based systems.
\newblock In {\em 2018 ACM/IEEE 45th Annual International Symposium on Computer Architecture (ISCA)}, pages 726--738, 2018.

\bibitem{zhong2011applicationspecific}
Wei Zhong, Bei Yu, Song Chen, Takeshi Yoshimura, Sheqin Dong, and Satoshi Goto.
\newblock Application-specific network-on-chip synthesis: Cluster generation and network component insertion.
\newblock In {\em 2011 12th International Symposium on Quality Electronic Design}, pages 1--6, 2011.

\end{thebibliography}

\end{document}